\documentclass[letterpaper,12pt]{article}
\pdfoutput=1
\usepackage{jheppub}

\usepackage{tikz}
\usepackage[export]{adjustbox}
\usepackage{subcaption}
\usepackage{empheq}
\usepackage{amsmath}
\usepackage{braket}

\newcommand{\cZ}{\mathcal{Z}}
\newcommand{\e}{\epsilon}
\newcommand{\smallminus}{{\rm\rule[2.4pt]{6pt}{0.65pt}}}
\newcommand{\union}{\cup}
\newcommand{\sq}[1]{[#1]}

\newcommand{\RN}[1]{\textup{\uppercase\expandafter{\romannumeral#1}}}

\DeclareMathOperator{\Gr}{Gr}
\DeclareMathOperator{\vol}{vol}

\allowdisplaybreaks[1]

\title{
\large Weak Separation, Positivity and Extremal Yangian Invariants
}

\author{Luke Lippstreu,$^{1}$}
\author{Jorge Mago,$^{1}$}
\author{Marcus Spradlin$^{2}$}
\author{and Anastasia Volovich$^{1}$}

\affiliation{$^1$ Department of Physics, Brown University, Providence, RI 02912, USA}

\affiliation{$^2$ Department of Physics and Brown Theoretical Physics Center, Brown University, Providence, RI 02912, USA}

\abstract{
We classify all positive $n$-particle N${}^k$MHV Yangian invariants in $\mathcal{N}=4$ Yang-Mills theory with $n=5k$, which we call extremal because none exist for $n>5k$. We show that this problem is equivalent to that of enumerating plane cactus graphs with $k$ pentagons. We use the known solution of that problem to provide an exact expression for the number of cyclic classes of such invariants for any $k$, and a simple rule for writing them down explicitly. We provide an alternative (but equivalent) classification by showing that a product of $k$ five-brackets with disjoint sets of indices is a positive Yangian invariant if and only if the sets are all weakly separated.
}

\begin{document}

\maketitle

\section{Introduction}

$\mathcal{N}=4$ supersymmetric Yang-Mills theory (SYM)~\cite{Brink:1976bc} provides a remarkable playground for physical mathematics. The conformal~\cite{Sohnius:1981sn} and dual conformal symmetry~\cite{Drummond:2008vq} of the planar theory close into a Yangian symmetry algebra~\cite{Drummond:2009fd}. This Yangian symmetry can be understood as a manifestation of T-duality in the dual string theory~\cite{Berkovits:2008ic,Beisert:2008iq,Beisert:2009cs}.

Yangian invariants (see for example~\cite{Mason:2009qx, ArkaniHamed:2009vw, ArkaniHamed:2009dg, ArkaniHamed:2009sx, Drummond:2010uq, Ashok:2010ie, ArkaniHamed:2012nw, Drummond:2010qh}) are the basic building blocks for many amplitude-related quantities of interest: any tree-level amplitude can be written as a linear combination of Yangian invariants~\cite{Drummond:2008cr}, and any leading singularity of a loop-level integrand is a Yangian invariant~\cite{ArkaniHamed:2010kv}. Thus a complete understanding of all Yangian invariant functions, and their properties, is of considerable interest.

Following~\cite{ArkaniHamed:2012nw,Arkani-Hamed:2017tmz} another key theme which has been emerging in the study of SYM is the importance of positive geometry. Indeed all of the abovementioned scattering amplitude quantities can be expressed in terms of positive (or at least, non-negative) Yangian invariants (defined below). There are no positive $n$-particle N${}^k$MHV Yangian invariants for $n>5k$. The purpose of this paper is to provide a complete classification of all positive Yangian invariants with $n=5k$, which we call \emph{extremal}.

Every extremal Yangian invariant is a product of five-brackets (defined in~\cite{Mason:2009qx}) but the converse is far from true; the restriction of positivity drastically restricts which products of five-brackets are allowed. We show that the combinatorial problem of enumerating extremal Yangian invariants is precisely encoded in the counting of plane cactus graphs~\cite{harary1953number} with $k$ pentagons, which was solved in~\cite{bona2000enumeration}. We are therefore able to provide an analytic expression for the number of cyclic equivalence classes of such invariants for any $k$, and a simple rule for writing them down explicitly.

We also find, and employ, an interesting direct connection between the condition for a product of five-brackets to be positive and a ``weak separation'' criterion~\cite{LZ,OPS} closely related to those that have appeared in other positive geometry problems. Specifically, we prove that a product of $k$ five-brackets with disjoint sets of indices corresponds to a positive Yangian if and only if the sets are all weakly separated.

The paper is organized as follows. In Sec.~\ref{Review} we review the relevant aspects of the positive Grassmannian integral and its connection to Yangian invariants. In Sec.~\ref{Building} we show that every extremal Yangian invariant is a product of five-brackets, and express the positivity criterion on such products as a combinatorial problem. This problem is solved in Sec.~\ref{Cacti} in terms of cactus graphs. In Sec.~\ref{Weakseparation} we introduce a notion of weak separation and explain its connection to positivity. We then introduce ``weak separation graphs" which provide an alternative, but equivalent, graphical tool for enumerating all extremal Yangian invariants.

In an ancillary {\sc Mathematica} file we include a complete list of the 7561 cyclic classes of extremal Yangian invariants for $1 \le k \le 7$.

\section{Yangian Invariants}\label{Review}

We begin by reviewing from~\cite{ArkaniHamed:2012nw} a few aspects of the construction of N${}^k$MHV Yangian invariants from Grassmannian integrals~\cite{ArkaniHamed:2009dn} that will be most relevant for our analysis. The basic objects of study are $n$-point \emph{on-shell functions}, which are residues of the top form on the $\Gr(k,n)$ momentum twistor Grassmannian~\cite{Mason:2009qx,ArkaniHamed:2009vw} expressible as
\begin{equation}\label{yangiandef}
f_\sigma(\mathcal{Z}) = \underset{C\subset\Gamma_\sigma}{\oint}\frac{d^{k\times n}C}{\vol{GL(k)}}\frac{\delta^{4k|4k}(C\cdot\cZ)}{M_1 M_2 \cdots M_n}\,.
\end{equation}
Here $\Gamma_\sigma$ is some $d$-dimensional cell in $\Gr(k,n)$, $\cZ_a^A = (Z_a^A, \eta_a^A)$ are momentum supertwistors~\cite{Hodges:2009hk} whose bosonic components $Z_a^A$ comprise $n$ homogeneous coordinates on $\mathbb{P}^3$, and $M_i = \det(C_i\, C_{i+1}\, \cdots C_{i{+}k})$ are (ordered, adjacent) maximal minors of $C$. Cells of dimension $d=4k$ are of special interest because the $4k$ integrals over the parameters of $C(\alpha_1,...,\alpha_{4k})\subset\Gamma_\sigma$ are entirely localized by the $4k$ bosonic delta functions, providing a non-trivial function of $\mathcal{Z}$ as long as $C \cdot Z = 0$ admits solutions for generic $Z$. The number of such solutions is called the intersection number $\Gamma^4(C)$ of the cell.

We will say that a quantity of the form~(\ref{yangiandef}) is an $n$-particle N${}^k$MHV \emph{positive Yangian invariant}, denoted $Y^{(k)}$, if $C$ is a $4k$-dimensional \emph{positroid} cell (which means that all ordered maximal minors of $C$ are non-negative), $C$ has no zero columns, and $\Gamma^4(C) > 0$. We have adopted a very narrow definition in order to focus precisely on the quantities of interest here, but we note that most of the literature uses a somewhat looser definition; in particular the earlier literature on Yangian invariants cited above, prior to~\cite{ArkaniHamed:2012nw}, did not require $C$ to be positive. Some aspects of the significance of positivity will play a crucial role in Sec.~\ref{Weakseparation}. The added requirement that $C$ has no zero columns (which was used in~\cite{ArkaniHamed:2012nw}) means that our positive $n$-particle Yangian invariants automatically have full functional dependence on all $n$ momentum twistors.

The simplest non-trivial Yangian invariant is the unique 5-particle NMHV invariant obtained from the cell
\begin{equation}\label{nmhvcmatrix}
C(\alpha_1,\ldots,\alpha_4) = \begin{pmatrix}
1 & \alpha_1 & \alpha_2 & \alpha_3 & \alpha_4
\end{pmatrix}.
\end{equation}
Note that here and in all following examples, we use parameterizations that render $C$-matrices non-negative when all $\alpha_i$ are positive. For this $C$-matrix~(\ref{yangiandef}) evaluates to the
\emph{five-bracket}~\cite{Mason:2009qx} (originally presented as an ``$R$-invariant'' in~\cite{Drummond:2008vq})
\begin{align}\label{fivebracket}
Y^{(1)} = [1,2,3,4,5] \equiv \frac{\delta^{(4)} (\langle 1\,2\,3\,4 \rangle \chi_5^A + \text{cyclic} ) }{\langle 1 \, 2 \, 3\,4 \rangle \langle 2 \, 3 \, 4 \,5 \rangle \langle 3 \, 4 \, 5 \, 1 \rangle \langle 4 \,5 \,1 \, 2 \rangle \langle 5 \, 1 \, 2 \, 3 \rangle}
\end{align}
where $\langle a\,b\,c\,d \rangle \equiv \epsilon_{ABCD} Z_a^A Z_b^B Z_c^C Z_d^D$ are the $SL(4)$-invariant Pl\"ucker coordinates. Note that the five-bracket is fully antisymmetric under the exchange of any two of its indices.

For any finite $k$ and $n$ the number of positive $n$-particle N${}^k$MHV Yangian invariants is finite. In the next section we review the fact that there are no invariants with $n > 5k$, so we will call those satisfying $n = 5 k$ \emph{extremal}. There we will also see that all extremal Yangian invariants are products of five-brackets with disjoint indices. For example, consider the 12-dimensional cell in $\Gr(3,15)$ with
\begin{equation}
C_1=
\begin{pmatrix}
 1 & \alpha_{1} & \alpha_{2} & \alpha_{3} & \alpha_4 & 0 & 0 & 0 & 0 & 0 & 0 & 0 & 0 & 0 & 0 \\
 0 & 0 & 0 & 0 & 0 & 1 & \alpha_5 & \alpha_6 & \alpha_7 & \alpha_8 & 0 & 0 & 0 & 0 & 0 \\
 0 & 0 & 0 & 0 & 0 & 0 & 0 & 0 & 0 & 0 & 1 & \alpha_9 & \alpha_{10} & \alpha_{11} & \alpha_{12} \\
\end{pmatrix},
\end{equation}
in which case~(\ref{yangiandef}) evaluates to
\begin{equation}\label{k3Yangian1}
Y^{(3)}_{1}=\sq{1,2,3,4,5}\sq{6,7,8,9,10}\sq{11,12,13,14,15}\,.
\end{equation}
We will shortly see that every $\Gr(3,15)$ Yangian invariant is cyclically equivalent (with respect to the $\mathbb{Z}_n$ symmetry that shifts $\mathcal{Z}_a \to \mathcal{Z}_{a+1\,{\rm mod}\,n}$) to~(\ref{k3Yangian1}) or one of
\begin{equation}\label{k3Yangian2}
\begin{aligned}
Y^{(3)}_{2}&=\sq{1,2,3,4,5}\sq{15,6,7,8,9}\sq{10,11,12,13,14}\,,\\
Y^{(3)}_{3}&=\sq{1,2,3,4,5}\sq{14,15,6,7,8}\sq{9,10,11,12,13}\,,
\end{aligned}
\end{equation}
which are obtained respectively from
\begin{equation}
\begin{aligned}
C_2 &=\begin{pmatrix}
 1 & \alpha_{1} & \alpha_{2} & \alpha_{3} & \alpha_4 & 0 & 0 & 0 & 0 & 0 & 0 & 0 & 0 & 0 & 0 \\
 0 & 0 & 0 & 0 & 0 & 1 & \alpha_5 & \alpha_6 & \alpha_7 & 0 & 0 & 0 & 0 & 0 & -\alpha_{12} \\
 0 & 0 & 0 & 0 & 0 & 0 & 0 & 0 & 0 & 1 & \alpha_8 & \alpha_{9} & \alpha_{10} & \alpha_{11} & 0 \\
\end{pmatrix}, \\
C_3 &= \begin{pmatrix}
 1 & \alpha_{1} & \alpha_{2} & \alpha_{3} & \alpha_4 & 0 & 0 & 0 & 0 & 0 & 0 & 0 & 0 & 0 & 0 \\
 0 & 0 & 0 & 0 & 0 & 1 & \alpha_5 & \alpha_6 & 0 & 0 & 0 & 0 & 0 & -\alpha_{11} & -\alpha_{12} \\
 0 & 0 & 0 & 0 & 0 & 0 & 0 & 0 & 1 & \alpha_7 & \alpha_{8} & \alpha_{9} & \alpha_{10} & 0 & 0 \\
\end{pmatrix}.
\end{aligned}
\end{equation}
For $\Gr(4,20)$ we will see that there are a total of 17 cyclic classes.

The classification of positive Yangian invariants for general $k$ and $n$ is discussed in Chapter~12 of~\cite{ArkaniHamed:2012nw}, where the number of cyclic classes is tabulated for $k \le 4$ in Table~3. We note however typos in the $(3,15)$ and $(4,20)$ entries, which were both underreported as 1 instead of 3 and 17. These missing Yangian invariants are precisely of the extremal type that are the focus of this paper. In the next section we describe the general construction of this interesting class of Yangian invariants.

\section{Building Extremal Yangian Invariants}\label{Building}

Using the lexicographic BCFW-bridge construction~\cite{ArkaniHamed:2012nw}, in order to build a $C$-matrix for some $4k$-dimensional cell one begins with a $k \times n$ matrix that has precisely $k$ non-zero entries equal to 1, one in each row and each column. Then one performs $4k$ ``shifts'' to populate $C$ with $4k$ parameters $\alpha_i$. Since we do not consider $C$ matrices with vanishing columns, we cannot have $n > 5 k$, and for $n=5k$ each of the $4k$ shifts must be used to populate precisely one of the $4k$ columns that are initially zero. Requiring $\Gamma^4(C)>0$, i.e.~demanding that there is always a solution to $C \cdot Z = 0$ for generic $Z \in \Gr(4,5k)$, requires that exactly four $\alpha$'s appear in each row. This means that $C\cdot Z=0$ imposes linear dependencies between $k$ disjoint subsets of 5 momentum twistors each. Therefore there are no Yangian invariants with $n>5 k$, and all invariants with $n=5k$ are products of five-brackets with disjoint indices.

In other words, every extremal Yangian invariant has the form
\begin{equation}\label{Ykgen}
\sq{1,2,3,4,5}\sq{6,7,8,9,10}\cdots\sq{n{-}4,n{-}3,n{-}2,n{-}1,n}
\end{equation}
up to some $S_n$ permutation of the particle labels. However, not all $n!$ permutations are allowed. Due to the disjoint nature of the linear dependencies, the on-shell diagram~\cite{ArkaniHamed:2012nw} associated to each such invariant will be the union of $k$ disconnected 5-particle on-shell diagrams, each involving only the five external lines whose indices appear in one of the five-brackets. The criterion for determining which permutations are valid is that the resulting on-shell diagram must be planar. In Sec.~\ref{Weakseparation} we indicate that this restriction to planar graphs is equivalent to restricting to the positive region of the Grassmannian.

\begin{figure}[t]
    \centering
    \begin{subfigure}{0.3\textwidth}
        \centering
        \includegraphics[width=0.8\textwidth]{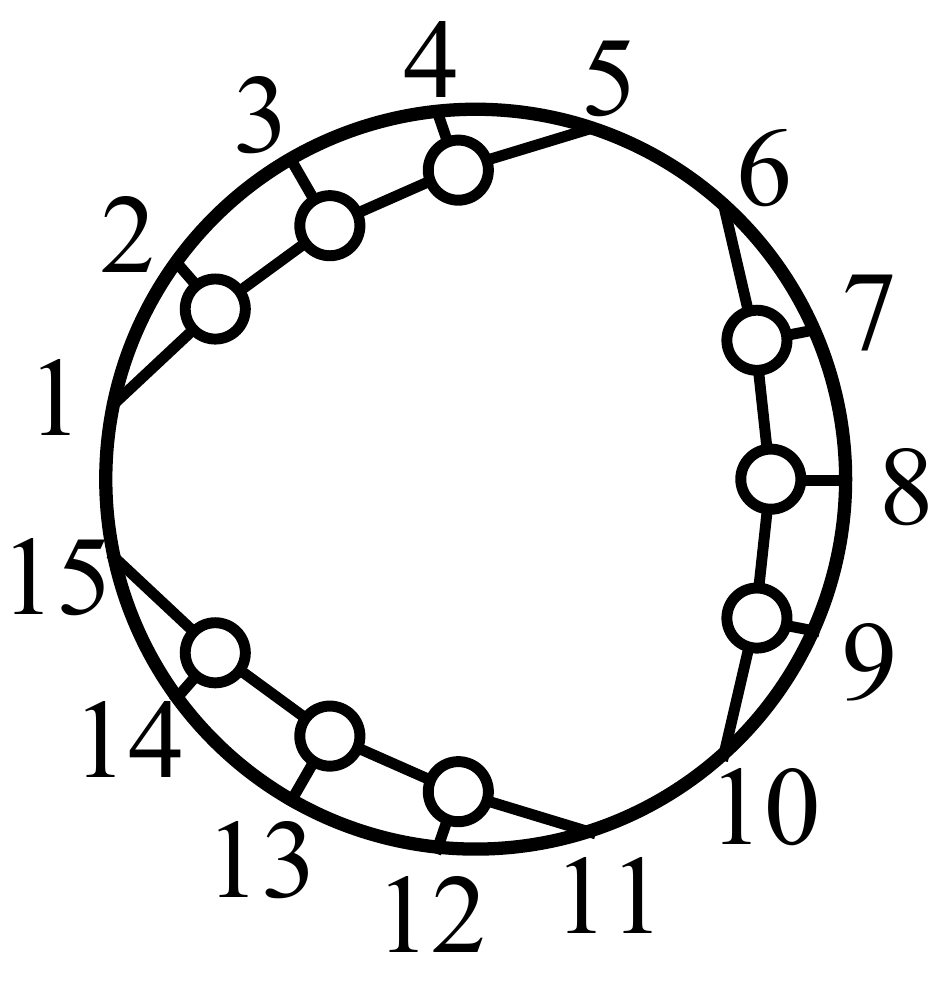}
        \caption*{$Y^{(3)}_{1}$}
        \label{Y3k1}
    \end{subfigure}
    \begin{subfigure}{0.3\textwidth}
        \centering
        \includegraphics[width=0.8\textwidth]{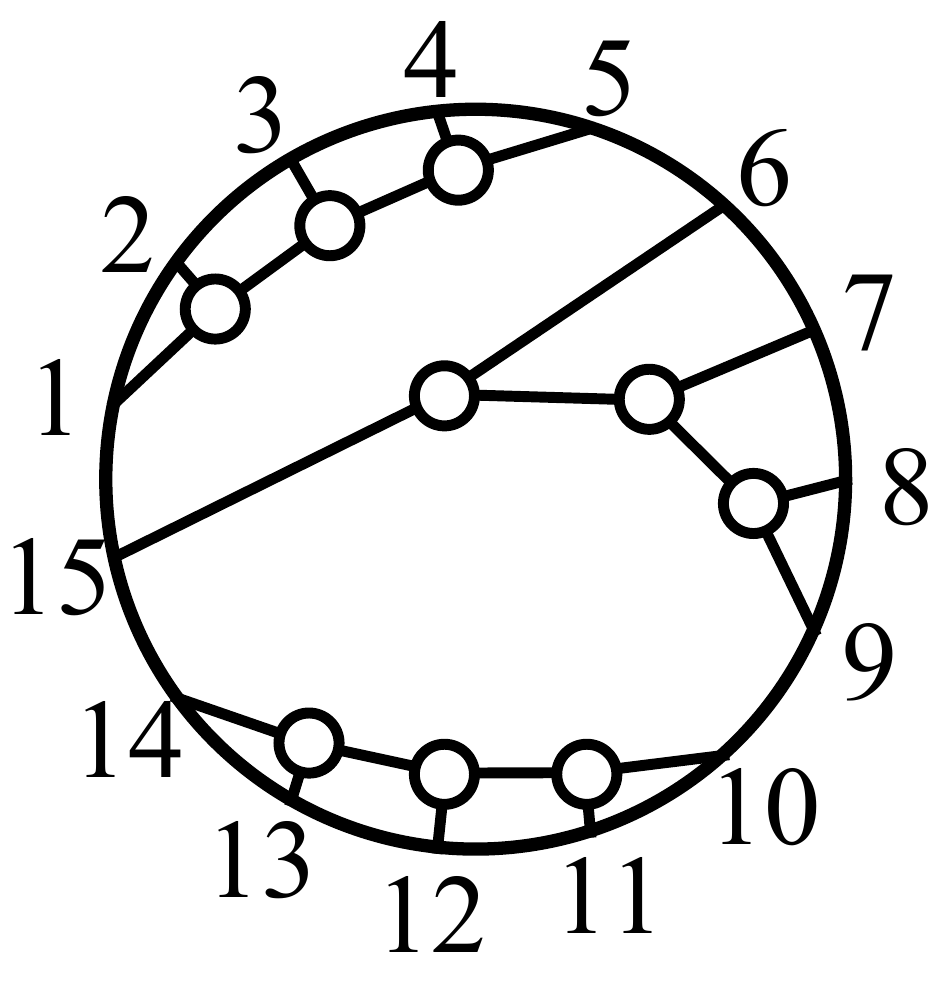}
        \caption*{$Y^{(3)}_{2}$}
        \label{Y3k2}
    \end{subfigure}
    \begin{subfigure}{0.3\textwidth}
        \centering
        \includegraphics[width=0.8\textwidth]{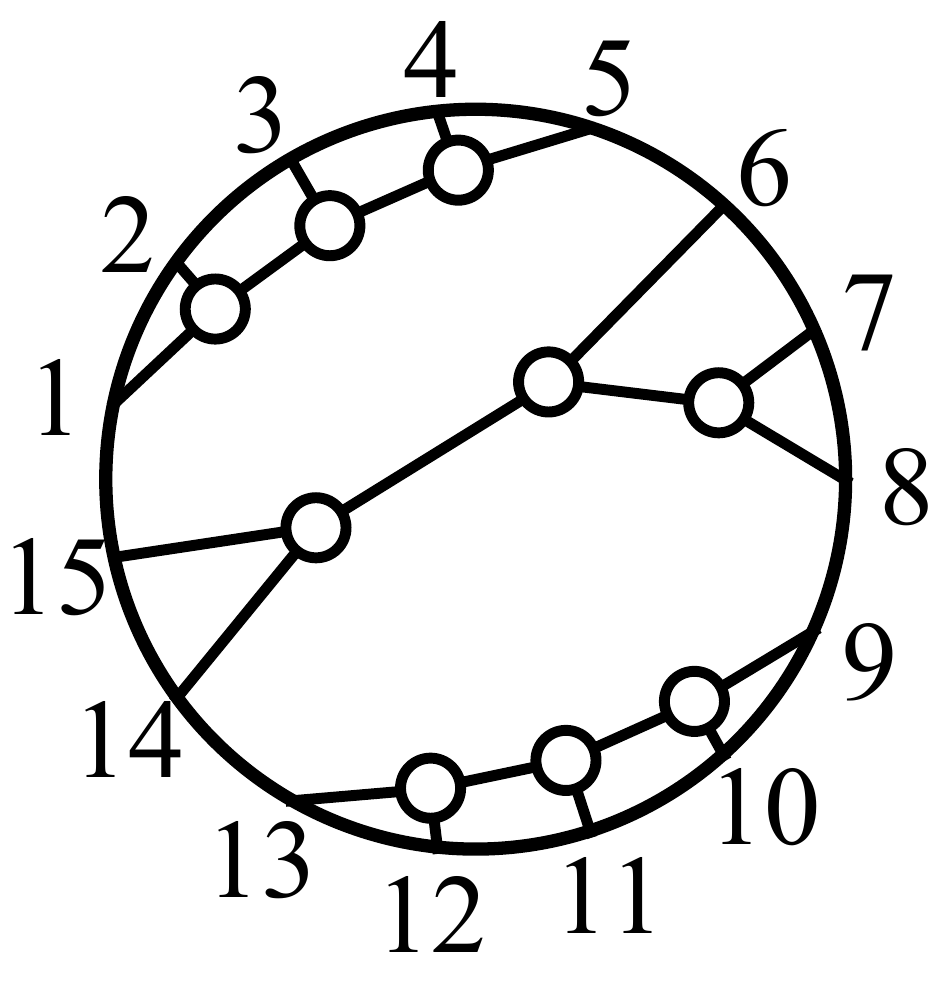}
        \caption*{$Y^{(3)}_{3}$}
        \label{Y3k3}
    \end{subfigure}
\caption{On-shell diagrams corresponding to the Yangian invariants (\ref{k3Yangian1}) and (\ref{k3Yangian2}).}
    \label{Y3k}
\end{figure}

For example, in the left panel of Fig.~\ref{Y3k} we have used~\cite{Bourjaily:2012gy} to draw the on-shell diagram corresponding to $Y_1^{(3)}$. Most permutations of the $n$ particle labels render this diagram non-planar, but a few do not. For example, note that $Y^{(3)}_2$ can be obtained from $Y^{(3)}_1$ by applying the cyclic permutation $\sigma: 15 \to 14 \to \cdots \to 6 \to 15$ to the indices of the last two five-brackets in~(\ref{k3Yangian1}), and applying the same $\sigma$ to the diagram on the left of Fig.~\ref{Y3k} produces the one in the middle. Applying $\sigma$ again turns $Y^{(3)}_2$ into $Y^{(3)}_3$ while transforming the diagram in the middle to the one on the right. Applying $\sigma$ a third time yields an on-shell diagram that evaluates to $Y^{(3)}_3$ again (due to the symmetry of the five-bracket), applying a fourth time gives one that evaluates to $Y^{(3)}_2$, and applying $\sigma$ a fifth time brings us back to $Y^{(3)}_1$.

It is clear that a graph of the type under consideration can be planar only if at least two five-brackets involve only cyclically adjacent indices. (This explains why every $\Gr(2,10)$ Yangian invariant is cyclically equivalent to $\sq{1,2,3,4,5} \sq{6,7,8,9,10}$.) As the examples in Fig.~\ref{Y3k} illustrate, for $k>2$ each of the other $k{-}2$ five-brackets can involve indices associated to a component of the graph that stretches from one gap between the two privileged five-brackets to the other. The precise ordering of particle number labels within each five-bracket is inconsequential due to the symmetries of the five-bracket; the Yangian invariant associated to such an on-shell diagram depends (up to an overall sign) only on how the indices $\set{1,\ldots,5k}$ are split into $k$ disjoint subsets as in~(\ref{Ykgen}).

In conclusion, the problem of enumerating all (cyclic classes of) extremal Yangian invariants reduces to the combinatorial problem of enumerating (cyclic classes of) planar on-shell graphs comprised of $k$ disconnected 5-particle on-shell diagrams, of the type shown in Fig.~\ref{Y3k}. This problem is equivalent to the problem of enumerating plane 5-gonal cactus graphs, to which we now turn our attention in the next section.

\section{Counting Cacti}\label{Cacti}

\emph{Cactus graphs} (or \emph{cacti}) were first defined in~\cite{harary1953number} under the name of ``Husimi trees". They are connected simple graphs in which every edge lies in a single elementary cycle. A cactus is called $m$-gonal if every elementary cycle is an $m$-gon. Finally, a plane $m$-gonal cactus is an $m$-gonal cactus drawn on the plane such that no edges cross and each edge is adjacent to the unbounded region. The following are examples of unlabeled plane 5-gonal cacti with 3 and 4 pentagons respectively:
\begin{align*}
        \begin{tikzpicture}[scale=0.5]
        \begin{scope}[every node/.style={circle,draw,inner sep=1pt, minimum size=1mm,fill}]
        \node (a) at (0.80338, 1.09902) {};
        \node (b) at (0., 0.592432) {};
        \node (c) at (0.345673, 0.) {};
        \node (d) at (1.18244, 0.44823) {};
        \node (e) at (1.94806, 1.32933) {};
        \node (f) at (2.31957, 2.43523) {};
        \node (g) at (2.28184, 3.38417) {};
        \node (h) at (1.5959, 3.38223) {};
        \node (i) at (1.56664, 2.43255) {};
        \node (j) at (2.72129, 0.454204) {};
        \node (k) at (3.56233, 0.0126173) {};
        \node (l) at (3.90213, 0.608273) {};
        \node (m) at (3.09472, 1.10792) {};
        \end{scope}
        \begin{scope}
        \path [-] (e) edge node {} (a);
        \path [-] (a) edge node {} (b);
        \path [-] (b) edge node {} (c);
        \path [-] (c) edge node {} (d);
        \path [-] (d) edge node {} (e);
        \path [-] (e) edge node {} (j);
        \path [-] (j) edge node {} (k);
        \path [-] (k) edge node {} (l);
        \path [-] (l) edge node {} (m);
        \path [-] (m) edge node {} (e);
        \path [-] (e) edge node {} (f);
        \path [-] (f) edge node {} (g);
        \path [-] (g) edge node {} (h);
        \path [-] (h) edge node {} (i);
        \path [-] (i) edge node {} (e);
        \end{scope}
        \end{tikzpicture}
\qquad\qquad\qquad
        \begin{tikzpicture}[scale=0.5]
        \begin{scope}[every node/.style={circle, draw, inner sep=0pt, minimum size=1mm, fill}]
        \node (a) at (1.15394, 2.76871) {};
        \node (b) at (0.497981, 3.45308) {};
        \node (c) at (0.0566464, 3.03183) {};
        \node (d) at (0.711002, 2.34513) {};
        \node (e) at (1.72633, 1.72546) {};
        \node (f) at (0.683865, 1.15167) {};
        \node (g) at (0., 0.495433) {};
        \node (h) at (0.422994, 0.0552169) {};
        \node (i) at (1.10764, 0.709457) {};
        \node (j) at (2.29939, 0.683229) {};
        \node (k) at (2.95597, 0.) {};
        \node (l) at (3.39723, 0.421049) {};
        \node (m) at (2.74212, 1.10631) {};
        \node (n) at (2.34583, 2.74056) {};
        \node (o) at (3.03215, 3.39462) {};
        \node (p) at (3.4547, 2.95412) {};
        \node (q) at (2.76903, 2.29831) {};
        \end{scope}
        \begin{scope}
        \path [-] (e) edge node {} (a);
        \path [-] (a) edge node {} (b);
        \path [-] (b) edge node {} (c);
        \path [-] (c) edge node {} (d);
        \path [-] (d) edge node {} (e);
        \path [-] (e) edge node {} (n);
        \path [-] (n) edge node {} (o);
        \path [-] (o) edge node {} (p);
        \path [-] (p) edge node {} (q);
        \path [-] (q) edge node {} (e);
        \path [-] (e) edge node {} (j);
        \path [-] (j) edge node {} (k);
        \path [-] (k) edge node {} (l);
        \path [-] (l) edge node {} (m);
        \path [-] (m) edge node {} (e);
        \path [-] (e) edge node {} (f);
        \path [-] (f) edge node {} (g);
        \path [-] (g) edge node {} (h);
        \path [-] (h) edge node {} (i);
        \path [-] (i) edge node {} (e);
        \end{scope}
        \end{tikzpicture}
\end{align*}

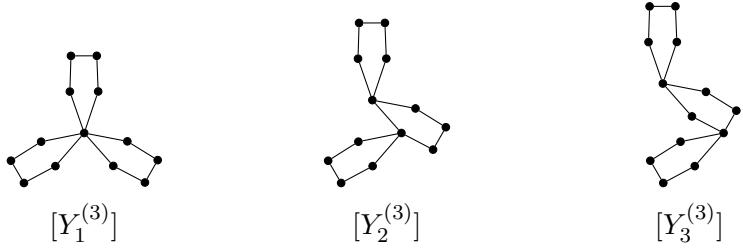
\begin{figure}[t]
\centering
        \begin{subfigure}[b]{0.25\linewidth}
	\centering
        \begin{tikzpicture}[scale=0.5]
        \begin{scope}[every node/.style={circle,draw,inner sep=1pt, minimum size=1mm,fill}]
        \node (a) at (0.80338, 1.09902) {};
        \node (b) at (0., 0.592432) {};
        \node (c) at (0.345673, 0.) {};
        \node (d) at (1.18244, 0.44823) {};
        \node (e) at (1.94806, 1.32933) {};
        \node (f) at (2.31957, 2.43523) {};
        \node (g) at (2.28184, 3.38417) {};
        \node (h) at (1.5959, 3.38223) {};
        \node (i) at (1.56664, 2.43255) {};
        \node (j) at (2.72129, 0.454204) {};
        \node (k) at (3.56233, 0.0126173) {};
        \node (l) at (3.90213, 0.608273) {};
        \node (m) at (3.09472, 1.10792) {};
        \end{scope}
        \begin{scope}
        \path [-] (e) edge node {} (a);
        \path [-] (a) edge node {} (b);
        \path [-] (b) edge node {} (c);
        \path [-] (c) edge node {} (d);
        \path [-] (d) edge node {} (e);
        \path [-] (e) edge node {} (j);
        \path [-] (j) edge node {} (k);
        \path [-] (k) edge node {} (l);
        \path [-] (l) edge node {} (m);
        \path [-] (m) edge node {} (e);
        \path [-] (e) edge node {} (f);
        \path [-] (f) edge node {} (g);
        \path [-] (g) edge node {} (h);
        \path [-] (h) edge node {} (i);
        \path [-] (i) edge node {} (e);
        \end{scope}
        \end{tikzpicture}
	\caption*{$[Y^{(3)}_{1}]$}
	\end{subfigure}
        \begin{subfigure}[b]{0.25\textwidth}
	\centering
        \begin{tikzpicture}[scale=0.5]
        \begin{scope}[every node/.style={circle,draw,inner sep=1pt, minimum size=1mm,fill}]
        \node (a) at (1.57661, 0.223896) {};
        \node (b) at (0.773232, -0.282696) {};
        \node (c) at (1.11891, -0.875128) {};
        \node (d) at (1.95567, -0.426898) {};
        \node (e) at (1.94806, 1.32933) {};
        \node (f) at (2.31957, 2.43523) {};
        \node (g) at (2.28184, 3.38417) {};
        \node (h) at (1.5959, 3.38223) {};
        \node (i) at (1.56664, 2.43255) {};
        \node (j) at (2.72129, 0.454204) {};
        \node (k) at (3.56233, 0.0126173) {};
        \node (l) at (3.90213, 0.608273) {};
        \node (m) at (3.09472, 1.10792) {};
        \end{scope}
        \begin{scope}
        \path [-] (j) edge node {} (a);
        \path [-] (a) edge node {} (b);
        \path [-] (b) edge node {} (c);
        \path [-] (c) edge node {} (d);
        \path [-] (d) edge node {} (j);
        \path [-] (e) edge node {} (j);
        \path [-] (j) edge node {} (k);
        \path [-] (k) edge node {} (l);
        \path [-] (l) edge node {} (m);
        \path [-] (m) edge node {} (e);
        \path [-] (e) edge node {} (f);
        \path [-] (f) edge node {} (g);
        \path [-] (g) edge node {} (h);
        \path [-] (h) edge node {} (i);
        \path [-] (i) edge node {} (e);
        \end{scope}
        \end{tikzpicture}
	\caption*{$[Y^{(3)}_{2}]$}
	\end{subfigure}
        \begin{subfigure}[b]{0.25\textwidth}
	\centering
        \begin{tikzpicture}[scale=0.5]
        \begin{scope}[every node/.style={circle,draw,inner sep=1pt, minimum size=1mm,fill}]
        \node (a) at (2.41765, -0.217691) {};
        \node (b) at (1.61427, -0.724283) {};
        \node (c) at (1.95995, -1.31672) {};
        \node (d) at (2.79671, -0.868485) {};
        \node (e) at (1.94806, 1.32933) {};
        \node (f) at (2.31957, 2.43523) {};
        \node (g) at (2.28184, 3.38417) {};
        \node (h) at (1.5959, 3.38223) {};
        \node (i) at (1.56664, 2.43255) {};
        \node (j) at (2.72129, 0.454204) {};
        \node (k) at (3.56233, 0.0126173) {};
        \node (l) at (3.90213, 0.608273) {};
        \node (m) at (3.09472, 1.10792) {};
        \end{scope}
        \begin{scope}
        \path [-] (k) edge node {} (a);
        \path [-] (a) edge node {} (b);
        \path [-] (b) edge node {} (c);
        \path [-] (c) edge node {} (d);
        \path [-] (d) edge node {} (k);
        \path [-] (e) edge node {} (j);
        \path [-] (j) edge node {} (k);
        \path [-] (k) edge node {} (l);
        \path [-] (l) edge node {} (m);
        \path [-] (m) edge node {} (e);
        \path [-] (e) edge node {} (f);
        \path [-] (f) edge node {} (g);
        \path [-] (g) edge node {} (h);
        \path [-] (h) edge node {} (i);
        \path [-] (i) edge node {} (e);
        \end{scope}
        \end{tikzpicture}
	\caption*{$[Y^{(3)}_{3}]$}
	\end{subfigure}
	\caption{The 3 plane 5-gonal cacti. Each corresponds to the indicated cyclic class $[Y_\alpha^{(3)}]$ of extremal Yangian invariants containing the element $Y_\alpha^{(3)}$ given in~(\ref{k3Yangian1}) or~(\ref{k3Yangian2}).}\label{3CactiDiagrams}.
\end{figure}

Cacti related by rotations on the plane are equivalent, but we can obtain cyclically inequivalent cacti by fixing one of the pentagons and then changing the shared vertices of two or more of the other pentagons. This change moves a pentagon around another one, and when performed five times in a row gives back the initial cactus. This process is similar to the one we described in the previous section for obtaining distinct extremal Yangians, where the choice of how many cyclic permutations to perform on $p$ five-brackets is in one-to-one correspondence with moving $p\smallminus1$ pentagons around another pentagon. For example, there are precisely three distinct plane 5-gonal cacti, and these relate to the three cyclic classes of $\Gr(3,15)$ Yangian invariants as indicated in Fig.~\ref{3CactiDiagrams}. Moving the lower left pentagon around the lower right one corresponds to applying the cyclic permutation $\sigma$ of the indices of the last two five-brackets in~(\ref{k3Yangian2}) that we considered in Sec.~\ref{Building}.

\begin{figure}[ht]
\centering
    \begin{subfigure}[b]{0.25\textwidth}
    \centering
        \begin{tikzpicture}[scale=0.50]
        \begin{scope}[every node/.style={circle, draw, inner sep=0pt, minimum size=1mm, fill}]
        \node (a) at (1.15394, 2.76871) {};
        \node (b) at (0.497981, 3.45308) {};
        \node (c) at (0.0566464, 3.03183) {};
        \node (d) at (0.711002, 2.34513) {};
        \node (e) at (1.72633, 1.72546) {};
        \node (f) at (0.683865, 1.15167) {};
        \node (g) at (0., 0.495433) {};
        \node (h) at (0.422994, 0.0552169) {};
        \node (i) at (1.10764, 0.709457) {};
        \node (j) at (2.29939, 0.683229) {};
        \node (k) at (2.95597, 0.) {};
        \node (l) at (3.39723, 0.421049) {};
        \node (m) at (2.74212, 1.10631) {};
        \node (n) at (2.34583, 2.74056) {};
        \node (o) at (3.03215, 3.39462) {};
        \node (p) at (3.4547, 2.95412) {};
        \node (q) at (2.76903, 2.29831) {};
        \end{scope}
        
        \begin{scope}
        \path [-] (e) edge node {} (a);
        \path [-] (a) edge node {} (b);
        \path [-] (b) edge node {} (c);
        \path [-] (c) edge node {} (d);
        \path [-] (d) edge node {} (e);
        \path [-] (e) edge node {} (n);
        \path [-] (n) edge node {} (o);
        \path [-] (o) edge node {} (p);
        \path [-] (p) edge node {} (q);
        \path [-] (q) edge node {} (e);
        \path [-] (e) edge node {} (j);
        \path [-] (j) edge node {} (k);
        \path [-] (k) edge node {} (l);
        \path [-] (l) edge node {} (m);
        \path [-] (m) edge node {} (e);
        \path [-] (e) edge node {} (f);
        \path [-] (f) edge node {} (g);
        \path [-] (g) edge node {} (h);
        \path [-] (h) edge node {} (i);
        \path [-] (i) edge node {} (e);
        \end{scope}
        \end{tikzpicture}
	\caption*{$[Y^{(4)}_{{1}}]$}
    \end{subfigure}
    \begin{subfigure}[b]{0.25\textwidth}
    \centering
        \begin{tikzpicture}[scale=0.50]
        \begin{scope}[every node/.style={circle,draw,inner sep=0pt, minimum size=1mm, fill}]
        \node (a) at (1.15394, 2.76871) {};
        \node (b) at (0.497981, 3.45308) {};
        \node (c) at (0.0566464, 3.03183) {};
        \node (d) at (0.711002, 2.34513) {};
        \node (e) at (1.72633, 1.72546) {};
        \node (f) at (1.25693, 0.109438) {};
        \node (g) at (0.573062, -0.546803) {};
        \node (h) at (0.996055, -0.987019) {};
        \node (i) at (1.6807, -0.332779) {};
        \node (j) at (2.29939, 0.683229) {};
        \node (k) at (2.95597, 0.) {};
        \node (l) at (3.39723, 0.421049) {};
        \node (m) at (2.74212, 1.10631) {};
        \node (n) at (2.34583, 2.74056) {};
        \node (o) at (3.03215, 3.39462) {};
        \node (p) at (3.4547, 2.95412) {};
        \node (q) at (2.76903, 2.29831) {};
        \end{scope}
        
        \begin{scope}
        \path [-] (e) edge node {} (a);
        \path [-] (a) edge node {} (b);
        \path [-] (b) edge node {} (c);
        \path [-] (c) edge node {} (d);
        \path [-] (d) edge node {} (e);
        \path [-] (e) edge node {} (n);
        \path [-] (n) edge node {} (o);
        \path [-] (o) edge node {} (p);
        \path [-] (p) edge node {} (q);
        \path [-] (q) edge node {} (e);
        \path [-] (e) edge node {} (j);
        \path [-] (j) edge node {} (k);
        \path [-] (k) edge node {} (l);
        \path [-] (l) edge node {} (m);
        \path [-] (m) edge node {} (e);
        \path [-] (j) edge node {} (f);
        \path [-] (f) edge node {} (g);
        \path [-] (g) edge node {} (h);
        \path [-] (h) edge node {} (i);
        \path [-] (i) edge node {} (j);
        \end{scope}
        \end{tikzpicture}
	\caption*{$[Y^{(4)}_{{2}}]$}
    \end{subfigure}
    \begin{subfigure}[b]{0.25\textwidth}
    \centering
        \begin{tikzpicture}[scale=0.50]
        \begin{scope}[every node/.style={circle,draw,inner sep=0pt, minimum size=1mm, fill}]
        \node (a) at (1.15394, 2.76871) {};
        \node (b) at (0.497981, 3.45308) {};
        \node (c) at (0.0566464, 3.03183) {};
        \node (d) at (0.711002, 2.34513) {};
        \node (e) at (1.72633, 1.72546) {};
        \node (f) at (1.91351, -0.573791) {};
        \node (g) at (1.22964, -1.23003) {};
        \node (h) at (1.65263, -1.67025) {};
        \node (i) at (2.33728, -1.01601) {};
        \node (j) at (2.29939, 0.683229) {};
        \node (k) at (2.95597, 0.) {};
        \node (l) at (3.39723, 0.421049) {};
        \node (m) at (2.74212, 1.10631) {};
        \node (n) at (2.34583, 2.74056) {};
        \node (o) at (3.03215, 3.39462) {};
        \node (p) at (3.4547, 2.95412) {};
        \node (q) at (2.76903, 2.29831) {};
        \end{scope}
        
        \begin{scope}
        \path [-] (e) edge node {} (a);
        \path [-] (a) edge node {} (b);
        \path [-] (b) edge node {} (c);
        \path [-] (c) edge node {} (d);
        \path [-] (d) edge node {} (e);
        \path [-] (e) edge node {} (n);
        \path [-] (n) edge node {} (o);
        \path [-] (o) edge node {} (p);
        \path [-] (p) edge node {} (q);
        \path [-] (q) edge node {} (e);
        \path [-] (e) edge node {} (j);
        \path [-] (j) edge node {} (k);
        \path [-] (k) edge node {} (l);
        \path [-] (l) edge node {} (m);
        \path [-] (m) edge node {} (e);
        \path [-] (k) edge node {} (f);
        \path [-] (f) edge node {} (g);
        \path [-] (g) edge node {} (h);
        \path [-] (h) edge node {} (i);
        \path [-] (i) edge node {} (k);
        \end{scope}
        \end{tikzpicture}
	\caption*{$[Y^{(4)}_{{3}}]$}
    \end{subfigure}
\vspace{0.07cm}
    \begin{subfigure}[b]{0.25\textwidth}
    \centering
        \begin{tikzpicture}[scale=0.50]
        \begin{scope}[every node/.style={circle,draw,inner sep=0pt, minimum size=1mm, fill}]
        \node (a) at (1.15394, 2.76871) {};
        \node (b) at (0.497981, 3.45308) {};
        \node (c) at (0.0566464, 3.03183) {};
        \node (d) at (0.711002, 2.34513) {};
        \node (e) at (1.72633, 1.72546) {};
        \node (f) at (4.01673, 1.43615) {};
        \node (g) at (4.70305, 2.09021) {};
        \node (h) at (5.1256, 1.64971) {};
        \node (i) at (4.43993, 0.993899) {};
        \node (j) at (2.29939, 0.683229) {};
        \node (k) at (2.95597, 0.) {};
        \node (l) at (3.39723, 0.421049) {};
        \node (m) at (2.74212, 1.10631) {};
        \node (n) at (2.34583, 2.74056) {};
        \node (o) at (3.03215, 3.39462) {};
        \node (p) at (3.4547, 2.95412) {};
        \node (q) at (2.76903, 2.29831) {};
        \end{scope}
        
        \begin{scope}
        \path [-] (e) edge node {} (a);
        \path [-] (a) edge node {} (b);
        \path [-] (b) edge node {} (c);
        \path [-] (c) edge node {} (d);
        \path [-] (d) edge node {} (e);
        \path [-] (e) edge node {} (n);
        \path [-] (n) edge node {} (o);
        \path [-] (o) edge node {} (p);
        \path [-] (p) edge node {} (q);
        \path [-] (q) edge node {} (e);
        \path [-] (e) edge node {} (j);
        \path [-] (j) edge node {} (k);
        \path [-] (k) edge node {} (l);
        \path [-] (l) edge node {} (m);
        \path [-] (m) edge node {} (e);
        \path [-] (l) edge node {} (f);
        \path [-] (f) edge node {} (g);
        \path [-] (g) edge node {} (h);
        \path [-] (h) edge node {} (i);
        \path [-] (i) edge node {} (l);
        \end{scope}
        \end{tikzpicture}
	\caption*{$[Y^{(4)}_{{4}}]$}
    \end{subfigure}
    \begin{subfigure}[b]{0.25\textwidth}
    \centering
        \begin{tikzpicture}[scale=0.50]
        \begin{scope}[every node/.style={circle,draw,inner sep=0pt, minimum size=1mm, fill}]
        \node (a) at (1.15394, 2.76871) {};
        \node (b) at (0.497981, 3.45308) {};
        \node (c) at (0.0566464, 3.03183) {};
        \node (d) at (0.711002, 2.34513) {};
        \node (e) at (1.72633, 1.72546) {};
        \node (f) at (3.36162, 2.12141) {};
        \node (g) at (4.04794, 2.77547) {};
        \node (h) at (4.47049, 2.33497) {};
        \node (i) at (3.78482, 1.67916) {};
        \node (j) at (2.29939, 0.683229) {};
        \node (k) at (2.95597, 0.) {};
        \node (l) at (3.39723, 0.421049) {};
        \node (m) at (2.74212, 1.10631) {};
        \node (n) at (2.34583, 2.74056) {};
        \node (o) at (3.03215, 3.39462) {};
        \node (p) at (3.4547, 2.95412) {};
        \node (q) at (2.76903, 2.29831) {};
        \end{scope}
        
        \begin{scope}
        \path [-] (e) edge node {} (a);
        \path [-] (a) edge node {} (b);
        \path [-] (b) edge node {} (c);
        \path [-] (c) edge node {} (d);
        \path [-] (d) edge node {} (e);
        \path [-] (e) edge node {} (n);
        \path [-] (n) edge node {} (o);
        \path [-] (o) edge node {} (p);
        \path [-] (p) edge node {} (q);
        \path [-] (q) edge node {} (e);
        \path [-] (e) edge node {} (j);
        \path [-] (j) edge node {} (k);
        \path [-] (k) edge node {} (l);
        \path [-] (l) edge node {} (m);
        \path [-] (m) edge node {} (e);
        \path [-] (m) edge node {} (f);
        \path [-] (f) edge node {} (g);
        \path [-] (g) edge node {} (h);
        \path [-] (h) edge node {} (i);
        \path [-] (i) edge node {} (m);
        \end{scope}
        \end{tikzpicture}
	\caption*{$[Y^{(4)}_{{5}}]$}
    \end{subfigure}
    \begin{subfigure}[b]{0.25\textwidth}
    \centering
        \begin{tikzpicture}[scale=0.50]
        \begin{scope}[every node/.style={circle,draw,inner sep=0pt, minimum size=1mm, fill}]
        \node (a) at (1.15394, 2.76871) {};
        \node (b) at (0.497981, 3.45308) {};
        \node (c) at (0.0566464, 3.03183) {};
        \node (d) at (0.711002, 2.34513) {};
        \node (e) at (1.72633, 1.72546) {};
        \node (f) at (2.29963, 0.682284) {};
        \node (g) at (1.61576, 0.0260427) {};
        \node (h) at (2.03876, -0.414173) {};
        \node (i) at (2.7234, 0.240067) {};
        \node (j) at (3.34209, 1.25607) {};
        \node (k) at (3.99867, 0.572846) {};
        \node (l) at (4.43993, 0.993895) {};
        \node (m) at (3.78482, 1.67916) {};
        \node (n) at (2.34583, 2.74056) {};
        \node (o) at (3.03215, 3.39462) {};
        \node (p) at (3.4547, 2.95412) {};
        \node (q) at (2.76903, 2.29831) {};
        \end{scope}
        
        \begin{scope}
        \path [-] (e) edge node {} (a);
        \path [-] (a) edge node {} (b);
        \path [-] (b) edge node {} (c);
        \path [-] (c) edge node {} (d);
        \path [-] (d) edge node {} (e);
        \path [-] (e) edge node {} (n);
        \path [-] (n) edge node {} (o);
        \path [-] (o) edge node {} (p);
        \path [-] (p) edge node {} (q);
        \path [-] (q) edge node {} (e);
        \path [-] (q) edge node {} (j);
        \path [-] (j) edge node {} (k);
        \path [-] (k) edge node {} (l);
        \path [-] (l) edge node {} (m);
        \path [-] (m) edge node {} (q);
        \path [-] (j) edge node {} (f);
        \path [-] (f) edge node {} (g);
        \path [-] (g) edge node {} (h);
        \path [-] (h) edge node {} (i);
        \path [-] (i) edge node {} (j);
        \end{scope}
        \end{tikzpicture}
	\caption*{$[Y^{(4)}_{{6}}]$}
    \end{subfigure}
\vspace{0.07cm}
    \begin{subfigure}[b]{0.25\textwidth}
    \centering
        \begin{tikzpicture}[scale=0.50]
        \begin{scope}[every node/.style={circle,draw,inner sep=0pt, minimum size=1mm, fill}]
        \node (a) at (1.15394, 2.76871) {};
        \node (b) at (0.497981, 3.45308) {};
        \node (c) at (0.0566464, 3.03183) {};
        \node (d) at (0.711002, 2.34513) {};
        \node (e) at (1.72633, 1.72546) {};
        \node (f) at (2.95621, -0.00094) {};
        \node (g) at (2.27234, -0.657181) {};
        \node (h) at (2.69534, -1.0974) {};
        \node (i) at (3.37998, -0.443157) {};
        \node (j) at (3.34209, 1.25607) {};
        \node (k) at (3.99867, 0.572846) {};
        \node (l) at (4.43993, 0.993895) {};
        \node (m) at (3.78482, 1.67916) {};
        \node (n) at (2.34583, 2.74056) {};
        \node (o) at (3.03215, 3.39462) {};
        \node (p) at (3.4547, 2.95412) {};
        \node (q) at (2.76903, 2.29831) {};
        \end{scope}
        
        \begin{scope}
        \path [-] (e) edge node {} (a);
        \path [-] (a) edge node {} (b);
        \path [-] (b) edge node {} (c);
        \path [-] (c) edge node {} (d);
        \path [-] (d) edge node {} (e);
        \path [-] (e) edge node {} (n);
        \path [-] (n) edge node {} (o);
        \path [-] (o) edge node {} (p);
        \path [-] (p) edge node {} (q);
        \path [-] (q) edge node {} (e);
        \path [-] (q) edge node {} (j);
        \path [-] (j) edge node {} (k);
        \path [-] (k) edge node {} (l);
        \path [-] (l) edge node {} (m);
        \path [-] (m) edge node {} (q);
        \path [-] (k) edge node {} (f);
        \path [-] (f) edge node {} (g);
        \path [-] (g) edge node {} (h);
        \path [-] (h) edge node {} (i);
        \path [-] (i) edge node {} (k);
        \end{scope}
        \end{tikzpicture}
	\caption*{$[Y^{(4)}_{{7}}]$}
    \end{subfigure}
    \begin{subfigure}[b]{0.25\textwidth}
    \centering
        \begin{tikzpicture}[scale=0.50]
        \begin{scope}[every node/.style={circle,draw,inner sep=0pt, minimum size=1mm, fill}]
        \node (a) at (1.15394, 2.76871) {};
        \node (b) at (0.497981, 3.45308) {};
        \node (c) at (0.0566464, 3.03183) {};
        \node (d) at (0.711002, 2.34513) {};
        \node (e) at (1.72633, 1.72546) {};
        \node (f) at (5.06193, 2.00833) {};
        \node (g) at (5.74949, 2.6607) {};
        \node (h) at (6.1694, 2.21754) {};
        \node (i) at (5.4838, 1.56431) {};
        \node (j) at (3.34209, 1.25607) {};
        \node (k) at (3.99867, 0.572846) {};
        \node (l) at (4.43993, 0.993895) {};
        \node (m) at (3.78482, 1.67916) {};
        \node (n) at (2.34583, 2.74056) {};
        \node (o) at (3.03215, 3.39462) {};
        \node (p) at (3.4547, 2.95412) {};
        \node (q) at (2.76903, 2.29831) {};
        \end{scope}
        
        \begin{scope}
        \path [-] (e) edge node {} (a);
        \path [-] (a) edge node {} (b);
        \path [-] (b) edge node {} (c);
        \path [-] (c) edge node {} (d);
        \path [-] (d) edge node {} (e);
        \path [-] (e) edge node {} (n);
        \path [-] (n) edge node {} (o);
        \path [-] (o) edge node {} (p);
        \path [-] (p) edge node {} (q);
        \path [-] (q) edge node {} (e);
        \path [-] (q) edge node {} (j);
        \path [-] (j) edge node {} (k);
        \path [-] (k) edge node {} (l);
        \path [-] (l) edge node {} (m);
        \path [-] (m) edge node {} (q);
        \path [-] (l) edge node {} (f);
        \path [-] (f) edge node {} (g);
        \path [-] (g) edge node {} (h);
        \path [-] (h) edge node {} (i);
        \path [-] (i) edge node {} (l);
        \end{scope}
        \end{tikzpicture}
	\caption*{$[Y^{(4)}_{{8}}]$}
    \end{subfigure}
    \begin{subfigure}[b]{0.25\textwidth}
    \centering
        \begin{tikzpicture}[scale=0.50]
        \begin{scope}[every node/.style={circle,draw,inner sep=0pt, minimum size=1mm, fill}]
        \node (a) at (1.15394, 2.76871) {};
        \node (b) at (0.497981, 3.45308) {};
        \node (c) at (0.0566464, 3.03183) {};
        \node (d) at (0.711002, 2.34513) {};
        \node (e) at (1.72633, 1.72546) {};
        \node (f) at (4.40682, 2.6936) {};
        \node (g) at (5.09438, 3.34596) {};
        \node (h) at (5.51429, 2.90281) {};
        \node (i) at (4.82869, 2.24958) {};
        \node (j) at (3.34209, 1.25607) {};
        \node (k) at (3.99867, 0.572846) {};
        \node (l) at (4.43993, 0.993895) {};
        \node (m) at (3.78482, 1.67916) {};
        \node (n) at (2.34583, 2.74056) {};
        \node (o) at (3.03215, 3.39462) {};
        \node (p) at (3.4547, 2.95412) {};
        \node (q) at (2.76903, 2.29831) {};
        \end{scope}
        
        \begin{scope}
        \path [-] (e) edge node {} (a);
        \path [-] (a) edge node {} (b);
        \path [-] (b) edge node {} (c);
        \path [-] (c) edge node {} (d);
        \path [-] (d) edge node {} (e);
        \path [-] (e) edge node {} (n);
        \path [-] (n) edge node {} (o);
        \path [-] (o) edge node {} (p);
        \path [-] (p) edge node {} (q);
        \path [-] (q) edge node {} (e);
        \path [-] (q) edge node {} (j);
        \path [-] (j) edge node {} (k);
        \path [-] (k) edge node {} (l);
        \path [-] (l) edge node {} (m);
        \path [-] (m) edge node {} (q);
        \path [-] (m) edge node {} (f);
        \path [-] (f) edge node {} (g);
        \path [-] (g) edge node {} (h);
        \path [-] (h) edge node {} (i);
        \path [-] (i) edge node {} (m);
        \end{scope}
        \end{tikzpicture}
	\caption*{$[Y^{(4)}_{{9}}]$}
    \end{subfigure}
\vspace{0.07cm}
    \begin{subfigure}[b]{0.25\textwidth}
    \centering
        \begin{tikzpicture}[scale=0.50]
        \begin{scope}[every node/.style={circle,draw,inner sep=0pt, minimum size=1mm, fill}]
        \node (a) at (1.15394, 2.76871) {};
        \node (b) at (0.497981, 3.45308) {};
        \node (c) at (0.0566464, 3.03183) {};
        \node (d) at (0.711002, 2.34513) {};
        \node (e) at (1.72633, 1.72546) {};
        \node (f) at (2.9853, 1.33809) {};
        \node (g) at (2.30143, 0.681853) {};
        \node (h) at (2.72443, 0.241637) {};
        \node (i) at (3.40907, 0.895877) {};
        \node (j) at (4.02776, 1.91188) {};
        \node (k) at (4.68434, 1.22866) {};
        \node (l) at (5.1256, 1.64971) {};
        \node (m) at (4.47049, 2.33497) {};
        \node (n) at (2.34583, 2.74056) {};
        \node (o) at (3.03215, 3.39462) {};
        \node (p) at (3.4547, 2.95412) {};
        \node (q) at (2.76903, 2.29831) {};
        \end{scope}
        
        \begin{scope}
        \path [-] (e) edge node {} (a);
        \path [-] (a) edge node {} (b);
        \path [-] (b) edge node {} (c);
        \path [-] (c) edge node {} (d);
        \path [-] (d) edge node {} (e);
        \path [-] (e) edge node {} (n);
        \path [-] (n) edge node {} (o);
        \path [-] (o) edge node {} (p);
        \path [-] (p) edge node {} (q);
        \path [-] (q) edge node {} (e);
        \path [-] (p) edge node {} (j);
        \path [-] (j) edge node {} (k);
        \path [-] (k) edge node {} (l);
        \path [-] (l) edge node {} (m);
        \path [-] (m) edge node {} (p);
        \path [-] (j) edge node {} (f);
        \path [-] (f) edge node {} (g);
        \path [-] (g) edge node {} (h);
        \path [-] (h) edge node {} (i);
        \path [-] (i) edge node {} (j);
        \end{scope}
        \end{tikzpicture}
	\caption*{$[Y^{(4)}_{{10}}]$}
    \end{subfigure}
    \begin{subfigure}[b]{0.25\textwidth}
    \centering
        \begin{tikzpicture}[scale=0.50]
        \begin{scope}[every node/.style={circle,draw,inner sep=0pt, minimum size=1mm, fill}]
        \node (a) at (1.15394, 2.76871) {};
        \node (b) at (0.497981, 3.45308) {};
        \node (c) at (0.0566464, 3.03183) {};
        \node (d) at (0.711002, 2.34513) {};
        \node (e) at (1.72633, 1.72546) {};
        \node (f) at (3.64188, 0.65487) {};
        \node (g) at (2.95801, -0.001367) {};
        \node (h) at (3.38101, -0.441583) {};
        \node (i) at (4.06565, 0.212657) {};
        \node (j) at (4.02776, 1.91188) {};
        \node (k) at (4.68434, 1.22866) {};
        \node (l) at (5.1256, 1.64971) {};
        \node (m) at (4.47049, 2.33497) {};
        \node (n) at (2.34583, 2.74056) {};
        \node (o) at (3.03215, 3.39462) {};
        \node (p) at (3.4547, 2.95412) {};
        \node (q) at (2.76903, 2.29831) {};
        \end{scope}
        
        \begin{scope}
        \path [-] (e) edge node {} (a);
        \path [-] (a) edge node {} (b);
        \path [-] (b) edge node {} (c);
        \path [-] (c) edge node {} (d);
        \path [-] (d) edge node {} (e);
        \path [-] (e) edge node {} (n);
        \path [-] (n) edge node {} (o);
        \path [-] (o) edge node {} (p);
        \path [-] (p) edge node {} (q);
        \path [-] (q) edge node {} (e);
        \path [-] (p) edge node {} (j);
        \path [-] (j) edge node {} (k);
        \path [-] (k) edge node {} (l);
        \path [-] (l) edge node {} (m);
        \path [-] (m) edge node {} (p);
        \path [-] (k) edge node {} (f);
        \path [-] (f) edge node {} (g);
        \path [-] (g) edge node {} (h);
        \path [-] (h) edge node {} (i);
        \path [-] (i) edge node {} (k);
        \end{scope}
        \end{tikzpicture}
	\caption*{$[Y^{(4)}_{{11}}]$}
    \end{subfigure}
    \begin{subfigure}[b]{0.25\textwidth}
    \centering
        \begin{tikzpicture}[scale=0.50]
        \begin{scope}[every node/.style={circle,draw,inner sep=0pt, minimum size=1mm, fill}]
        \node (a) at (1.15394, 2.76871) {};
        \node (b) at (0.497981, 3.45308) {};
        \node (c) at (0.0566464, 3.03183) {};
        \node (d) at (0.711002, 2.34513) {};
        \node (e) at (1.72633, 1.72546) {};
        \node (f) at (5.7476, 2.66414) {};
        \node (g) at (6.43515, 3.31652) {};
        \node (h) at (6.85507, 2.87336) {};
        \node (i) at (6.16947, 2.22012) {};
        \node (j) at (4.02776, 1.91188) {};
        \node (k) at (4.68434, 1.22866) {};
        \node (l) at (5.1256, 1.64971) {};
        \node (m) at (4.47049, 2.33497) {};
        \node (n) at (2.34583, 2.74056) {};
        \node (o) at (3.03215, 3.39462) {};
        \node (p) at (3.4547, 2.95412) {};
        \node (q) at (2.76903, 2.29831) {};
        \end{scope}
        
        \begin{scope}
        \path [-] (e) edge node {} (a);
        \path [-] (a) edge node {} (b);
        \path [-] (b) edge node {} (c);
        \path [-] (c) edge node {} (d);
        \path [-] (d) edge node {} (e);
        \path [-] (e) edge node {} (n);
        \path [-] (n) edge node {} (o);
        \path [-] (o) edge node {} (p);
        \path [-] (p) edge node {} (q);
        \path [-] (q) edge node {} (e);
        \path [-] (p) edge node {} (j);
        \path [-] (j) edge node {} (k);
        \path [-] (k) edge node {} (l);
        \path [-] (l) edge node {} (m);
        \path [-] (m) edge node {} (p);
        \path [-] (l) edge node {} (f);
        \path [-] (f) edge node {} (g);
        \path [-] (g) edge node {} (h);
        \path [-] (h) edge node {} (i);
        \path [-] (i) edge node {} (l);
        \end{scope}
        \end{tikzpicture}
	\caption*{$[Y^{(4)}_{{12}}]$}
    \end{subfigure}
\vspace{0.07cm}
    \begin{subfigure}[b]{0.25\textwidth}
    \centering
        \begin{tikzpicture}[scale=0.50]
        \begin{scope}[every node/.style={circle,draw,inner sep=0pt, minimum size=1mm, fill}]
        \node (a) at (1.15394, 2.76871) {};
        \node (b) at (0.497981, 3.45308) {};
        \node (c) at (0.0566464, 3.03183) {};
        \node (d) at (0.711002, 2.34513) {};
        \node (e) at (1.72633, 1.72546) {};
        \node (f) at (4.64791, 2.92523) {};
        \node (g) at (5.30449, 2.24201) {};
        \node (h) at (5.74575, 2.66306) {};
        \node (i) at (5.09064, 3.34832) {};
        \node (j) at (3.65165, 4.40972) {};
        \node (k) at (4.33797, 5.06378) {};
        \node (l) at (4.76052, 4.62328) {};
        \node (m) at (4.07485, 3.96747) {};
        \node (n) at (2.34583, 2.74056) {};
        \node (o) at (3.03215, 3.39462) {};
        \node (p) at (3.4547, 2.95412) {};
        \node (q) at (2.76903, 2.29831) {};
        \end{scope}
        
        \begin{scope}
        \path [-] (e) edge node {} (a);
        \path [-] (a) edge node {} (b);
        \path [-] (b) edge node {} (c);
        \path [-] (c) edge node {} (d);
        \path [-] (d) edge node {} (e);
        \path [-] (e) edge node {} (n);
        \path [-] (n) edge node {} (o);
        \path [-] (o) edge node {} (p);
        \path [-] (p) edge node {} (q);
        \path [-] (q) edge node {} (e);
        \path [-] (o) edge node {} (j);
        \path [-] (j) edge node {} (k);
        \path [-] (k) edge node {} (l);
        \path [-] (l) edge node {} (m);
        \path [-] (m) edge node {} (o);
        \path [-] (m) edge node {} (f);
        \path [-] (f) edge node {} (g);
        \path [-] (g) edge node {} (h);
        \path [-] (h) edge node {} (i);
        \path [-] (i) edge node {} (m);
        \end{scope}
        \end{tikzpicture}
	\caption*{$[Y^{(4)}_{{13}}]$}
    \end{subfigure}
    \begin{subfigure}[b]{0.25\textwidth}
    \centering
        \begin{tikzpicture}[scale=0.50]
        \begin{scope}[every node/.style={circle,draw,inner sep=0pt, minimum size=1mm, fill}]
        \node (a) at (1.15394, 2.76871) {};
        \node (b) at (0.497981, 3.45308) {};
        \node (c) at (0.0566464, 3.03183) {};
        \node (d) at (0.711002, 2.34513) {};
        \node (e) at (1.72633, 1.72546) {};
        \node (f) at (5.33358, 3.58104) {};
        \node (g) at (5.99016, 2.89782) {};
        \node (h) at (6.43142, 3.31887) {};
        \node (i) at (5.77631, 4.00413) {};
        \node (j) at (3.65165, 4.40972) {};
        \node (k) at (4.33797, 5.06378) {};
        \node (l) at (4.76052, 4.62328) {};
        \node (m) at (4.07485, 3.96747) {};
        \node (n) at (2.34583, 2.74056) {};
        \node (o) at (3.03215, 3.39462) {};
        \node (p) at (3.4547, 2.95412) {};
        \node (q) at (2.76903, 2.29831) {};
        \end{scope}
        
        \begin{scope}
        \path [-] (e) edge node {} (a);
        \path [-] (a) edge node {} (b);
        \path [-] (b) edge node {} (c);
        \path [-] (c) edge node {} (d);
        \path [-] (d) edge node {} (e);
        \path [-] (e) edge node {} (n);
        \path [-] (n) edge node {} (o);
        \path [-] (o) edge node {} (p);
        \path [-] (p) edge node {} (q);
        \path [-] (q) edge node {} (e);
        \path [-] (o) edge node {} (j);
        \path [-] (j) edge node {} (k);
        \path [-] (k) edge node {} (l);
        \path [-] (l) edge node {} (m);
        \path [-] (m) edge node {} (o);
        \path [-] (l) edge node {} (f);
        \path [-] (f) edge node {} (g);
        \path [-] (g) edge node {} (h);
        \path [-] (h) edge node {} (i);
        \path [-] (i) edge node {} (l);
        \end{scope}
        \end{tikzpicture}
	\caption*{$[Y^{(4)}_{{14}}]$}
    \end{subfigure}
    \begin{subfigure}[b]{0.25\textwidth}
    \centering
        \begin{tikzpicture}[scale=0.50]
        \begin{scope}[every node/.style={circle,draw,inner sep=0pt, minimum size=1mm, fill}]
        \node (a) at (1.15394, 2.76871) {};
        \node (b) at (0.497981, 3.45308) {};
        \node (c) at (0.0566464, 3.03183) {};
        \node (d) at (0.711002, 2.34513) {};
        \node (e) at (1.72633, 1.72546) {};
        \node (f) at (2.39294, 4.79891) {};
        \node (g) at (3.07926, 5.45297) {};
        \node (h) at (3.50181, 5.01247) {};
        \node (i) at (2.81614, 4.35666) {};
        \node (j) at (1.77344, 3.78381) {};
        \node (k) at (1.11748, 4.46818) {};
        \node (l) at (0.676146, 4.04693) {};
        \node (m) at (1.3305, 3.36023) {};
        \node (n) at (2.34583, 2.74056) {};
        \node (o) at (3.03215, 3.39462) {};
        \node (p) at (3.4547, 2.95412) {};
        \node (q) at (2.76903, 2.29831) {};
        \end{scope}
        
        \begin{scope}
        \path [-] (e) edge node {} (a);
        \path [-] (a) edge node {} (b);
        \path [-] (b) edge node {} (c);
        \path [-] (c) edge node {} (d);
        \path [-] (d) edge node {} (e);
        \path [-] (e) edge node {} (n);
        \path [-] (n) edge node {} (o);
        \path [-] (o) edge node {} (p);
        \path [-] (p) edge node {} (q);
        \path [-] (q) edge node {} (e);
        \path [-] (n) edge node {} (j);
        \path [-] (j) edge node {} (k);
        \path [-] (k) edge node {} (l);
        \path [-] (l) edge node {} (m);
        \path [-] (m) edge node {} (n);
        \path [-] (j) edge node {} (f);
        \path [-] (f) edge node {} (g);
        \path [-] (g) edge node {} (h);
        \path [-] (h) edge node {} (i);
        \path [-] (i) edge node {} (j);
        \end{scope}
        \end{tikzpicture}
	\caption*{$[Y^{(4)}_{{15}}]$}
    \end{subfigure}
\vspace{0.07cm}
    \begin{subfigure}[b]{0.25\textwidth}
    \centering
        \begin{tikzpicture}[scale=0.50]
        \begin{scope}[every node/.style={circle,draw,inner sep=0pt, minimum size=1mm, fill}]
        \node (a) at (1.15394, 2.76871) {};
        \node (b) at (0.497981, 3.45308) {};
        \node (c) at (0.0566464, 3.03183) {};
        \node (d) at (0.711002, 2.34513) {};
        \node (e) at (1.72633, 1.72546) {};
        \node (f) at (1.25693, 0.109438) {};
        \node (g) at (0.573062, -0.546803) {};
        \node (h) at (0.996055, -0.987019) {};
        \node (i) at (1.6807, -0.332779) {};
        \node (j) at (2.29939, 0.683229) {};
        \node (k) at (2.95597, 0.) {};
        \node (l) at (3.39723, 0.421049) {};
        \node (m) at (2.74212, 1.10631) {};
        \node (n) at (3.36162, 2.12141) {};
        \node (o) at (4.04794, 2.77547) {};
        \node (p) at (4.47049, 2.33497) {};
        \node (q) at (3.78482, 1.67916) {};
        \end{scope}
        
        \begin{scope}
        \path [-] (e) edge node {} (a);
        \path [-] (a) edge node {} (b);
        \path [-] (b) edge node {} (c);
        \path [-] (c) edge node {} (d);
        \path [-] (d) edge node {} (e);
        \path [-] (m) edge node {} (n);
        \path [-] (n) edge node {} (o);
        \path [-] (o) edge node {} (p);
        \path [-] (p) edge node {} (q);
        \path [-] (q) edge node {} (m);
        \path [-] (e) edge node {} (j);
        \path [-] (j) edge node {} (k);
        \path [-] (k) edge node {} (l);
        \path [-] (l) edge node {} (m);
        \path [-] (m) edge node {} (e);
        \path [-] (j) edge node {} (f);
        \path [-] (f) edge node {} (g);
        \path [-] (g) edge node {} (h);
        \path [-] (h) edge node {} (i);
        \path [-] (i) edge node {} (j);
        \end{scope}
        
        \end{tikzpicture}
	\caption*{$[Y^{(4)}_{{16}}]$}
    \end{subfigure}
    \begin{subfigure}[b]{0.25\textwidth}
    \centering
        \begin{tikzpicture}[scale=0.50]
        \begin{scope}[every node/.style={circle,draw,inner sep=0pt, minimum size=1mm, fill}]
        \node (a) at (1.15394, 2.76871) {};
        \node (b) at (0.497981, 3.45308) {};
        \node (c) at (0.0566464, 3.03183) {};
        \node (d) at (0.711002, 2.34513) {};
        \node (e) at (1.72633, 1.72546) {};
        \node (f) at (1.25693, 0.109438) {};
        \node (g) at (0.573062, -0.546803) {};
        \node (h) at (0.996055, -0.987019) {};
        \node (i) at (1.6807, -0.332779) {};
        \node (j) at (2.29939, 0.683229) {};
        \node (k) at (2.95597, 0.) {};
        \node (l) at (3.39723, 0.421049) {};
        \node (m) at (2.74212, 1.10631) {};
        \node (n) at (4.01673, 1.43615) {};
        \node (o) at (4.70305, 2.09021) {};
        \node (p) at (5.1256, 1.64971) {};
        \node (q) at (4.43993, 0.993899) {};
        \end{scope}
        
        \begin{scope}
        \path [-] (e) edge node {} (a);
        \path [-] (a) edge node {} (b);
        \path [-] (b) edge node {} (c);
        \path [-] (c) edge node {} (d);
        \path [-] (d) edge node {} (e);
        \path [-] (l) edge node {} (n);
        \path [-] (n) edge node {} (o);
        \path [-] (o) edge node {} (p);
        \path [-] (p) edge node {} (q);
        \path [-] (q) edge node {} (l);
        \path [-] (e) edge node {} (j);
        \path [-] (j) edge node {} (k);
        \path [-] (k) edge node {} (l);
        \path [-] (l) edge node {} (m);
        \path [-] (m) edge node {} (e);
        \path [-] (j) edge node {} (f);
        \path [-] (f) edge node {} (g);
        \path [-] (g) edge node {} (h);
        \path [-] (h) edge node {} (i);
        \path [-] (i) edge node {} (j);
        \end{scope}
        \end{tikzpicture}
	\caption*{$[Y^{(4)}_{{17}}]$}
    \end{subfigure}
    \caption{The 17 plane 5-gonal cacti. Each corresponds to the indicated cyclic class $[Y_\alpha^{(4)}]$ of extremal Yangian invariants containing the element $Y_\alpha^{(4)}$ given in~(\ref{k4yangians}).}\label{5goncacti}
\end{figure}
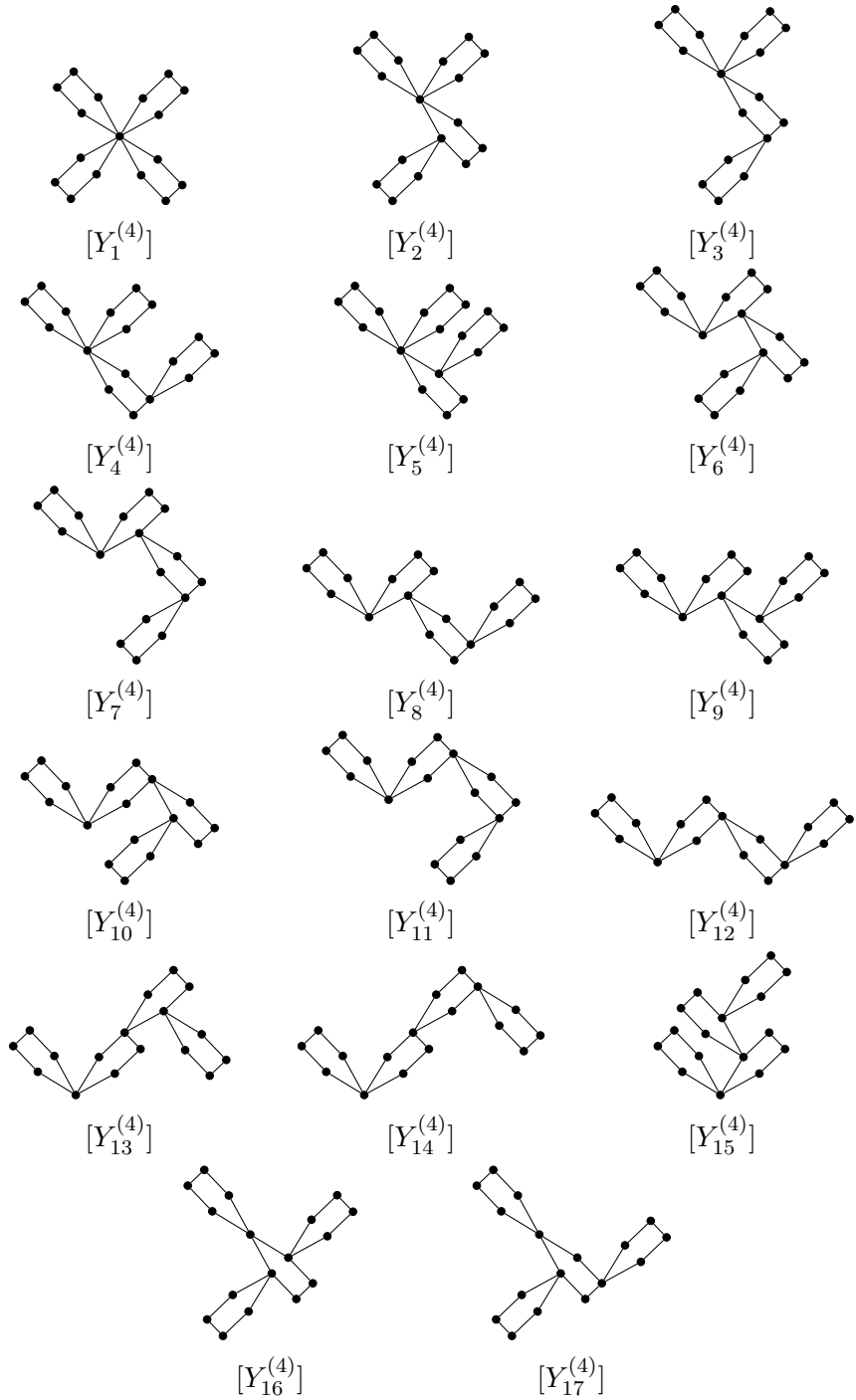

Similarly, there are 17 distinct plane 5-gonal cacti shown in Fig.~\ref{5goncacti}, and each is matched to a cyclic class of $\Gr(4,20)$ Yangian invariants, generated for example by
\begin{equation}\label{k4yangians}
\begin{aligned}
Y^{(4)}_{1}&=R_{1,2,4}R_{6,7,9}R_{11,12,14}R_{16,17,19}\,,&\quad Y^{(4)}_{2}&=R_{1,2,4}R_{6,7,9}R_{20,11,13}R_{15,16,18}\,,\\
Y^{(4)}_{3}&=R_{1,2,4}R_{6,7,9}R_{11,12,19}R_{14,15,17}\,,&\quad Y^{(4)}_{4}&=R_{1,2,4}R_{6,7,9}R_{18,19,11}R_{13,14,16}\,,\\
Y^{(4)}_{5}&=R_{1,2,4}R_{6,7,9}R_{11,17,19}R_{12,13,15}\,,&\quad
Y^{(4)}_{6}&=R_{1,2,4}R_{20,6,8}R_{19,10,12}R_{14,15,17}\,,\\
Y^{(4)}_{7}&=R_{1,2,4}R_{20,6,8}R_{10,11,18}R_{13,14,16}\,,&\quad
Y^{(4)}_{8}&=R_{1,2,4}R_{20,6,8}R_{19,10,17}R_{12,13,15}\,,\\
Y^{(4)}_{9}&=R_{1,2,4}R_{20,6,8}R_{10,16,18}R_{11,12,14}\,,&\quad
Y^{(4)}_{{10}}&=R_{1,2,4}R_{6,7,19}R_{18,9,11}R_{13,14,16}\,,\\
Y^{(4)}_{{11}}&=R_{1,2,4}R_{6,7,19}R_{9,10,17}R_{12,13,15}\,,&\quad
Y^{(4)}_{{12}}&=R_{1,2,4}R_{6,7,19}R_{18,9,16}R_{11,12,14}\,,\\
Y^{(4)}_{{13}}&=R_{1,2,4}R_{20,6,18}R_{17,8,10}R_{12,13,15}\,,&\quad
Y^{(4)}_{{14}}&=R_{1,2,4}R_{20,6,18}R_{8,9,16}R_{11,12,14}\,,\\
Y^{(4)}_{{15}}&=R_{1,2,4}R_{6,17,19}R_{16,7,9}R_{11,12,14}\,,&\quad
Y^{(4)}_{{16}}&=R_{1,2,4}R_{9,10,12}R_{15,16,18}\sq{6,7,8,14,20}\,,\\
Y^{(4)}_{{17}}&=R_{1,2,4}R_{20,6,13}R_{8,9,11}R_{15,16,18}\,,
\end{aligned}
\end{equation}
where we use $R_{i,j,k}=\sq{i,j,j{+}1,k,k{+}1}$ for notational efficiency.

The enumeration of several kinds of cacti was carried out~\cite{bona2000enumeration}. In particular, the number of unlabeled plane $m$-gonal cacti having $k$ polygons is given by
\begin{equation}\label{Htilde}
\tilde{\mathcal{H}}_k=\alpha_k+\beta_k-\gamma_k,
\end{equation}
in terms of
\begin{equation}
\begin{aligned}
\alpha_k&=\frac{1}{mk}\sum_{d|k}\phi\left(\frac{k}{d}\right)\binom{dm}{d},\\
\beta_k&=\frac{1}{mk}\sum_{d|(m,k-1)}\phi(d)\binom{km/d}{(k-1)/d},\\
\gamma_k&=\frac{1}{k(m-1)+1}\binom{mk}{k},
\end{aligned}
\end{equation}
where $\phi$ is Euler's totient function. We conclude that the number of cyclic classes of $\Gr(k,5k)$ Yangian invariants is then given by $\tilde{\mathcal{H}}_k$ with $m=5$ (see Tab.~\ref{tableofnumbers}).

\begin{table}[ht]
\centering
\begin{tabular}{cc}
$k$ & $\tilde{\mathcal{H}}_k$ \\ \hline
1   & 1                       \\
2   & 1                       \\
3   & 3                       \\
4   & 17                      \\
5   & 102                     \\
6   & 811                     \\
7   & 6626                    \\
8   & 58385                   \\
9   & 532251                  \\
10  & 5011934                 \\
11  & 48344880                \\
12  & 475982471               \\
13  & 4766639628              \\
14  & 48434621610             \\
15  & 498363430232
\end{tabular}
\caption{The number of cyclic classes of $\Gr(k,5k)$ Yangian invariants for $k \leq 15$, given by~(\ref{Htilde}). Also the number of plane 5-gonal cacti with $k$ pentagons~\cite{bona2000enumeration} (see also~\cite{OEIS}).}
\label{tableofnumbers}
\end{table}

\section{Weak Separation and Positivity}\label{Weakseparation}

In this section we demonstrate an alternative procedure for generating all positive extremal Yangian invariants. This method will have the added benefit of making it transparent that it is the positivity requirement which most strongly restricts the allowed form of products of five-brackets at $n=5k$.

As explained in Sec.~\ref{Building}, all extremal Yangian invariants are products of $k$ five-brackets involving disjoint particle numbers. In fact, \emph{any} product of five-brackets involving disjoint particle numbers is Yangian invariant, but not necessarily positive. A relevant way to see this is to use the result of~\cite{Drummond:2010uq} that the Grassmannian integral~(\ref{yangiandef}) is Yangian invariant for any $C$-matrix, and it is always possible to find an appropriate $C$-matrix such that~(\ref{yangiandef}) evaluates to any desired product of five-brackets. For example, in order to produce the product $[1,3,4,5,6][2,7,8,9,10]$ we could choose
\begin{equation}
  C=\begin{pmatrix}
  1&0&\beta_1&\beta_2&\beta_3&\beta_4&0&0&0&0\\
  0&1&0&0&0&0&\gamma_1&\gamma_2&\gamma_3&\gamma_4
  \end{pmatrix}\label{egmat}
\end{equation}
(the general rule is that the $r^{th}$ row refers to the $r^{th}$ five-bracket, the columns refer to the particle number, and we gauge fix the first non-zero entry in each row to 1). Evaluating the integral~(\ref{yangiandef}) for this $C$-matrix produces our product $[1,3,4,5,6][2,7,8,9,10]$, which is therefore Yangian invariant. From this example it should be clear that any product of five-brackets involving disjoint particle labels is Yangian invariant.

However, a brief examination of the matrix~(\ref{egmat}) reveals that this Yangian invariant is not \emph{positive}. That is, there is no nontrivial choice of domain for the  $\beta$ and $\gamma$ parameters which renders the matrix non-negative. To see this, examine the ordered minors $\braket{2,3}=-\beta_1$, $\braket{3,7}=\beta_1\gamma_1$ and $\braket{1,7}=\gamma_1$. Non-negativity of the first minor implies $\beta_1\le 0$, but then the second implies $\gamma_1 \le 0$, while the third requires $\gamma_1 \ge 0$, which is a contradiction unless $\gamma_1 = 0$ which we disallow because it gives a zero column. One must also check whether a row interchange alleviates the contradiction, and one finds that it does not.  Therefore our example is Yangian invariant but not positive.

It turns out that the necessary and sufficient condition for a disjoint product of five-brackets to have a non-negative $C$-matrix is that the five-brackets all be \textit{weakly separated}. The notion of weak separation was first introduced in \cite{LZ}. There they defined two $r$-element subsets $I$ and $J$ of the integers $\set{1,2,...,n} \equiv [n]$ to be \textit{weakly separated} if there exists a chord separating the sets $I\backslash J$ and $J\backslash I$ drawn on a circle (here $I\backslash J$ denotes the relative complement of $J$ with respect to $I$; $I\backslash J=\set{x: x\in I; x\notin J}$). In our application the sets we encounter will always be disjoint, so the necessity for taking the relative compliments evaporates and the  weak separation requirement becomes simply that there exist a chord separating the two sets $I$ and $J$ when drawn on a circle.  We now state this as our main theorem, which is a special case of a more general theorem proven in appendix~\ref{proofappendix}.

\bigskip
\noindent
   \textbf{Theorem:} A partition of $[5k]$ into $k$ subsets of length five corresponds to a \textit{non-negative} $C$-matrix if and only if the subsets are \textit{weakly separated}.

\bigskip

Using this theorem, we conclude that the set of positive extremal Yangian invariants is in perfect correspondence with the set of five-bracket partitions of $[n]$ for which each pair of five-brackets is weakly separated.

It should be mentioned that the relevance of the notion of weak separation to plabic graphs has been studied previously in~\cite{OPS}. There it was shown that maximally weakly separated collections in a positroid are in bijective correspondence with plabic graphs. However, there the maximally separated collections correspond to the face labels of the plabic graph, which is a completely different construction to the one we encounter. The concept of weak separation has also appeared in some recent papers~\cite{Golden:2019kks,MSSV} on the cluster structure of scattering amplitudes in SYM theory, since~\cite{OPS} showed that two Pl\"ucker coordinates $\langle I \rangle$, $\langle J \rangle$ are cluster adjacent~\cite{Drummond:2017ssj} if and only if $I$ and $J$ are weakly separated.

\begin{figure}
\centering
\begin{adjustbox}{minipage=\linewidth,scale=0.88}
  	\begin{subfigure}[h]{0.16\textwidth}
	\centering
			\begin{tikzpicture}
				\def \n {20}
				\def \radius {0.8}
				\draw circle(\radius)
				foreach\s in{1,...,\n}{
					(-360/\n*\s:-\radius)circle(.4pt)circle(.8pt)circle(1.2pt)
					node[anchor=-360/\n*\s](\s){${}$}
				};
				\draw[very thick](1) to [out=-15,in=-90] (5);
				
				\draw[very thick](6) to [out=-110,in=180] (10);
				\draw[very thick](11) to [out=160,in=90] (15);
			\end{tikzpicture}
			\caption*{$[Y_{1}^{(4)}]$}
  		\end{subfigure}
  	  	  	  	\begin{subfigure}[h]{0.16\textwidth}
  	  	  	  	\centering
  	  	  				\begin{tikzpicture}
  	  	  					\def \n {20}
  	  	  					\def \radius {0.8}
  	  	  					\draw circle(\radius)
  	  	  					foreach\s in{1,...,\n}{
  	  	  						(-360/\n*\s:-\radius)circle(.4pt)circle(.8pt)circle(1.2pt)
  	  	  						node[anchor=-360/\n*\s](\s){${}$}
  	  	  					};
  	  	  					\draw[very thick](1) to [out=-15,in=-90] (5);
  	  	  					
  	  	  					\draw[very thick](6) to [out=-110,in=180] (10);
  	  	  					\draw[very thick](15) to [out=80,in=20] (19);
  	  	  				\end{tikzpicture}
  	  	  				\caption*{$[Y_{2}^{(4)}]$}
  	  	  	  		\end{subfigure}
  	  	  	\begin{subfigure}[h]{0.16\textwidth}
  	  	  	\centering
  	  				\begin{tikzpicture}
  	  					\def \n {20}
  	  					\def \radius {0.8}
  	  					\draw circle(\radius)
  	  					foreach\s in{1,...,\n}{
  	  						(-360/\n*\s:-\radius)circle(.4pt)circle(.8pt)circle(1.2pt)
  	  						node[anchor=-360/\n*\s](\s){${}$}
  	  					};
  	  					\draw[very thick](1) to [out=-15,in=-90] (5);
  	  					
  	  					\draw[very thick](6) to [out=-110,in=180] (10);
  	  					\draw[very thick](14) to [out=100,in=40] (18);
  	  				\end{tikzpicture}
  	  				\caption*{$[Y_{3}^{(4)}]$}
  	  	  		\end{subfigure}
  	  	\begin{subfigure}[h]{0.16\textwidth}
  	  	\centering
  				\begin{tikzpicture}
  					\def \n {20}
  					\def \radius {0.8}
  					\draw circle(\radius)
  					foreach\s in{1,...,\n}{
  						(-360/\n*\s:-\radius)circle(.4pt)circle(.8pt)circle(1.2pt)
  						node[anchor=-360/\n*\s](\s){${}$}
  					};
  					\draw[very thick](1) to [out=-15,in=-90] (5);
  					
  					\draw[very thick](6) to [out=-110,in=180] (10);
  					\draw[very thick](13) to [out=120,in=60] (17);
  				\end{tikzpicture}
  				\caption*{$[Y_{4}^{(4)}]$}
  	  		\end{subfigure}
  	\begin{subfigure}[h]{0.16\textwidth}
  		\centering
			\begin{tikzpicture}
				\def \n {20}
				\def \radius {0.8}
				\draw circle(\radius)
				foreach\s in{1,...,\n}{
					(-360/\n*\s:-\radius)circle(.4pt)circle(.8pt)circle(1.2pt)
					node[anchor=-360/\n*\s](\s){${}$}
				};
				\draw[very thick](1) to [out=-15,in=-90] (5);
				
				\draw[very thick](6) to [out=-110,in=180] (10);
				\draw[very thick](12) to [out=140,in=70] (16);
			\end{tikzpicture}
			\caption*{$[Y_{5}^{(4)}]$}
  		\end{subfigure}
   	  	  	  	\begin{subfigure}[h]{0.16\textwidth}
   	  	  	  		\centering
   	  	  				\begin{tikzpicture}
   	  	  					\def \n {20}
   	  	  					\def \radius {0.8}
   	  	  					\draw circle(\radius)
   	  	  					foreach\s in{1,...,\n}{
   	  	  						(-360/\n*\s:-\radius)circle(.4pt)circle(.8pt)circle(1.2pt)
   	  	  						node[anchor=-360/\n*\s](\s){${}$}
   	  	  					};
   	  	  					\draw[very thick](1) to [out=-15,in=-90] (5);
   	  	  					
   	  	  					\draw[very thick](8) to [out=-150,in=150] (12);
   	  	  					\draw[very thick](16) to [out=70,in=-130] (7);
   	  	  				\end{tikzpicture}
   	  	  				\caption*{$[Y_{{6}}^{(4)}]$}
   	  	  	  		\end{subfigure}
   	  	  	  	\begin{subfigure}[h]{0.16\textwidth}
   	  	  	  		\centering
   	  	  				\begin{tikzpicture}
   	  	  					\def \n {20}
   	  	  					\def \radius {0.8}
   	  	  					\draw circle(\radius)
   	  	  					foreach\s in{1,...,\n}{
   	  	  						(-360/\n*\s:-\radius)circle(.4pt)circle(.8pt)circle(1.2pt)
   	  	  						node[anchor=-360/\n*\s](\s){${}$}
   	  	  					};
   	  	  					\draw[very thick](1) to [out=-15,in=-90] (5);
   	  	  					
   	  	  					\draw[very thick](9) to [out=-170,in=120] (13);
   	  	  					\draw[very thick](17) to [out=60,in=-150] (8);
   	  	  				\end{tikzpicture}
   	  	  				\caption*{$[Y_{7}^{(4)}]$}
   	  	  	  		\end{subfigure}
   	  	  	  	\begin{subfigure}[h]{0.16\textwidth}
   	  	  	  		\centering
   	  	  				\begin{tikzpicture}
   	  	  					\def \n {20}
   	  	  					\def \radius {0.8}
   	  	  					\draw circle(\radius)
   	  	  					foreach\s in{1,...,\n}{
   	  	  						(-360/\n*\s:-\radius)circle(.4pt)circle(.8pt)circle(1.2pt)
   	  	  						node[anchor=-360/\n*\s](\s){${}$}
   	  	  					};
   	  	  					\draw[very thick](1) to [out=-15,in=-90] (5);
   	  	  					
   	  	  					\draw[very thick](10) to [out=-180,in=110] (14);
   	  	  					\draw[very thick](18) to [out=40,in=-170] (9);
   	  	  				\end{tikzpicture}
   	  	  				\caption*{$[Y_{{8}}^{(4)}]$}
   	  	  	  		\end{subfigure}
   	  	  	  	\begin{subfigure}[h]{0.16\textwidth}
   	  	  	  		\centering
   	  	  				\begin{tikzpicture}
   	  	  					\def \n {20}
   	  	  					\def \radius {0.8}
   	  	  					\draw circle(\radius)
   	  	  					foreach\s in{1,...,\n}{
   	  	  						(-360/\n*\s:-\radius)circle(.4pt)circle(.8pt)circle(1.2pt)
   	  	  						node[anchor=-360/\n*\s](\s){${}$}
   	  	  					};
   	  	  					\draw[very thick](1) to [out=-15,in=-90] (5);
   	  	  					
   	  	  					\draw[very thick](11) to [out=-200,in=90] (15);
   	  	  					\draw[very thick](19) to [out=20,in=-180] (10);
   	  	  				\end{tikzpicture}
   	  	  				\caption*{$[Y_{9}^{(4)}]$}
   	  	  	  		\end{subfigure}
   	  	  	  	\begin{subfigure}[h]{0.16\textwidth}
   	  	  	  		\centering
   	  	  				\begin{tikzpicture}
   	  	  					\def \n {20}
   	  	  					\def \radius {0.8}
   	  	  					\draw circle(\radius)
   	  	  					foreach\s in{1,...,\n}{
   	  	  						(-360/\n*\s:-\radius)circle(.4pt)circle(.8pt)circle(1.2pt)
   	  	  						node[anchor=-360/\n*\s](\s){${}$}
   	  	  					};
   	  	  					\draw[very thick](1) to [out=-15,in=-90] (5);
   	  	  					
   	  	  					\draw[very thick](13) to [out=130,in=50] (17);
   	  	  					\draw[very thick](18) to [out=40,in=-170] (9);
   	  	  				\end{tikzpicture}
   	  	  				\caption*{$[Y_{{10}}^{(4)}]$}
   	  	  	  		\end{subfigure}
   	  	  	  	\begin{subfigure}[h]{0.16\textwidth}
   	  	  	  		\centering
   	  	  				\begin{tikzpicture}
   	  	  					\def \n {20}
   	  	  					\def \radius {0.8}
   	  	  					\draw circle(\radius)
   	  	  					foreach\s in{1,...,\n}{
   	  	  						(-360/\n*\s:-\radius)circle(.4pt)circle(.8pt)circle(1.2pt)
   	  	  						node[anchor=-360/\n*\s](\s){${}$}
   	  	  					};
   	  	  					\draw[very thick](1) to [out=-15,in=-90] (5);
   	  	  					
   	  	  					\draw[very thick](12) to [out=150,in=70] (16);
   	  	  					\draw[very thick](18) to [out=40,in=-170] (9);
   	  	  				\end{tikzpicture}
   	  	  				\caption*{$[Y_{{11}}^{(4)}]$}
   	  	  	  		\end{subfigure}
    	  	  	  	\begin{subfigure}[h]{0.16\textwidth}
    	  	  	  		\centering
    	  	  				\begin{tikzpicture}
    	  	  					\def \n {20}
    	  	  					\def \radius {0.8}
    	  	  					\draw circle(\radius)
    	  	  					foreach\s in{1,...,\n}{
    	  	  						(-360/\n*\s:-\radius)circle(.4pt)circle(.8pt)circle(1.2pt)
    	  	  						node[anchor=-360/\n*\s](\s){${}$}
    	  	  					};
    	  	  					\draw[very thick](1) to [out=-15,in=-90] (5);
    	  	  					
    	  	  					\draw[very thick](11) to [out=160,in=90] (15);
    	  	  					\draw[very thick](18) to [out=40,in=-170] (9);
    	  	  				\end{tikzpicture}
    	  	  				\caption*{$[Y_{{12}}^{(4)}]$}
    	  	  	  		\end{subfigure}
   	  	  	  	\begin{subfigure}[h]{0.16\textwidth}
   	  	  	  		\centering
   	  	  				\begin{tikzpicture}
   	  	  					\def \n {20}
   	  	  					\def \radius {0.8}
   	  	  					\draw circle(\radius)
   	  	  					foreach\s in{1,...,\n}{
   	  	  						(-360/\n*\s:-\radius)circle(.4pt)circle(.8pt)circle(1.2pt)
   	  	  						node[anchor=-360/\n*\s](\s){${}$}
   	  	  					};
   	  	  					\draw[very thick](1) to [out=-15,in=-90] (5);
   	  	  					
   	  	  					\draw[very thick](10) to [out=-180,in=110] (14);
   	  	  					\draw[very thick](16) to [out=70,in=-130] (7);
   	  	  				\end{tikzpicture}
   	  	  				\caption*{$[Y_{{13}}^{(4)}]$}
   	  	  	  		\end{subfigure}
   	  	  	  	\begin{subfigure}[h]{0.16\textwidth}
   	  	  	  		\centering
   	  	  				\begin{tikzpicture}
   	  	  					\def \n {20}
   	  	  					\def \radius {0.8}
   	  	  					\draw circle(\radius)
   	  	  					foreach\s in{1,...,\n}{
   	  	  						(-360/\n*\s:-\radius)circle(.4pt)circle(.8pt)circle(1.2pt)
   	  	  						node[anchor=-360/\n*\s](\s){${}$}
   	  	  					};
   	  	  					\draw[very thick](1) to [out=-15,in=-90] (5);
   	  	  					
   	  	  					\draw[very thick](11) to [out=160,in=90] (15);
   	  	  					\draw[very thick](17) to [out=50,in=-150] (8);
   	  	  				\end{tikzpicture}
   	  	  				\caption*{$[Y_{{14}}^{(4)}]$}
   	  	  	  		\end{subfigure}
   	  	  	  	\begin{subfigure}[h]{0.16\textwidth}
   	  	  	  		\centering
   	  	  				\begin{tikzpicture}
   	  	  					\def \n {20}
   	  	  					\def \radius {0.8}
   	  	  					\draw circle(\radius)
   	  	  					foreach\s in{1,...,\n}{
   	  	  						(-360/\n*\s:-\radius)circle(.4pt)circle(.8pt)circle(1.2pt)
   	  	  						node[anchor=-360/\n*\s](\s){${}$}
   	  	  					};
   	  	  					\draw[very thick](1) to [out=-15,in=-90] (5);
   	  	  					
   	  	  					\draw[very thick](11) to [out=170,in=90] (15);
   	  	  					\draw[very thick](16) to [out=70,in=-130] (7);
   	  	  				\end{tikzpicture}
   	  	  				\caption*{$[Y_{{15}}^{(4)}]$}
   	  	  	  		\end{subfigure}
  	  	  	  	\begin{subfigure}[h]{0.16\textwidth}
  	  	  	  		\centering
  	  	  				\begin{tikzpicture}
  	  	  					\def \n {20}
  	  	  					\def \radius {0.8}
  	  	  					\draw circle(\radius)
  	  	  					foreach\s in{1,...,\n}{
  	  	  						(-360/\n*\s:-\radius)circle(.4pt)circle(.8pt)circle(1.2pt)
  	  	  						node[anchor=-360/\n*\s](\s){${}$}
  	  	  					};
  	  	  					\draw[very thick](1) to [out=-15,in=-90] (5);
  	  	  					
  	  	  					\draw[very thick](7) to [out=-130,in=170] (11);
  	  	  					\draw[very thick](15) to [out=80,in=20] (19);
  	  	  				\end{tikzpicture}
  	  	  				\caption*{$[Y_{{16}}^{(4)}]$}
  	  	  	  		\end{subfigure}
  	  	  	  	\begin{subfigure}[h]{0.16\textwidth}
  	  	  	  		\centering
  	  	  				\begin{tikzpicture}
  	  	  					\def \n {20}
  	  	  					\def \radius {0.8}
  	  	  					\draw circle(\radius)
  	  	  					foreach\s in{1,...,\n}{
  	  	  						(-360/\n*\s:-\radius)circle(.4pt)circle(.8pt)circle(1.2pt)
  	  	  						node[anchor=-360/\n*\s](\s){${}$}
  	  	  					};
  	  	  					\draw[very thick](1) to [out=-15,in=-90] (5);
  	  	  					
  	  	  					\draw[very thick](8) to [out=-150,in=150] (12);
  	  	  					\draw[very thick](15) to [out=80,in=20] (19);
  	  	  				\end{tikzpicture}
  	  	  				\caption*{$ [Y_{{17}}^{(4)}]$}
  	  	  	  		\end{subfigure}
  	  	  	  	\centering
  	  	  	  	\end{adjustbox}
   	  	  	  	\centering
   	  	  	  	\caption{The 17 cyclically distinct weak separation graphs for $n=20,k=4$. Each corresponds to the indicated cyclic class $[Y_\alpha^{(4)}]$ of extremal Yangian invariants containing the element $Y_\alpha^{(4)}$ given in~(\ref{k4yangians}).}\label{k=4}
\end{figure}
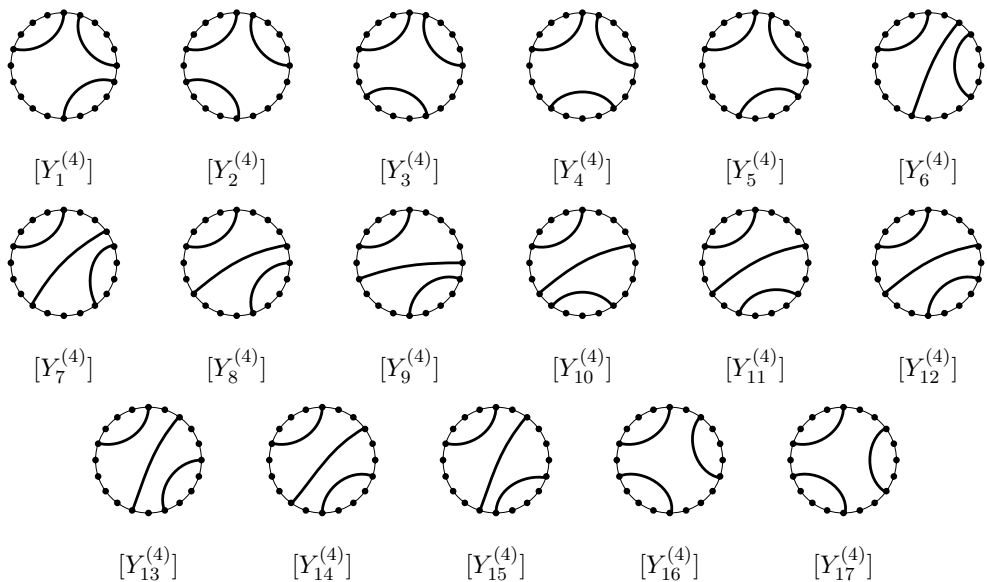

A convenient graphical tool for finding all cyclically distinct weakly separated partitions of $[n]$ into $k$ disjoint subsets of length $5$ is to draw $k{-}1$ chords on a circle that separate the integers into $k$ such sets. We call these \emph{weak separation graphs}. The cyclically distinct weak separation graphs for $k=3,n=15$ and $k=4,n=20$ are displayed in Figs.~\ref{k=3} and~\ref{k=4}, respectively.

Before concluding this section we remark that the above theorem is trivial if one observes the structure of on-shell graphs corresponding to the product of $k$ five-brackets involving disjoint particle labels. For example, examining the on-shell graphs given in Fig.~\ref{Y3k} it is clear that all such positive Yangian invariants are obtained by gluing multiple copies of the primitive $[a,b,c,d,e]$ on-shell graph onto the circle whilst preserving planarity. One can then see the rather straightforward correspondence between on-shell graphs and weak separation graphs. One can then make use of the fact that on-shell graphs always correspond to positive regions of the Grassmannian~\cite{ArkaniHamed:2012nw}. This then implies that any weakly separated product of five-brackets involving disjoint particle labels corresponds to a non-negative $C$-matrix. In the appendix we show that the converse is also true. One can then use the interchangeability of weak separation and planarity in this case to make the statement that the appearance of positive geometry is, at least as far as the extremal Yangians are concerned, a consequence of having restricted to the planar limit of the theory. That aspects of positive geometry arise more generally in planar theories has already been emphasized in~\cite{ArkaniHamed:2012nw}.

\begin{figure}
     	\centering
     	\begin{subfigure}[h]{0.29\textwidth}
	\centering
   			\begin{tikzpicture}
   				\def \n {15}
   				\def \radius {1.2}
   				\draw circle(\radius)
   				foreach\s in{1,...,\n}{
   					(-360/\n*\s:-\radius)circle(.4pt)circle(.8pt)circle(1.2pt)
   					node[anchor=-360/\n*\s](\s){}
   				};
   				\draw[very thick](1) to [out=-15,in=-120] (5);
   				
   				\draw[very thick](6) to [out=-140,in=120] (10);
   			\end{tikzpicture}
   			\caption*{$[Y_{1}^{(3)}]$}
     		\end{subfigure}
     	\begin{subfigure}[h]{0.29\textwidth}
	\centering
   			\begin{tikzpicture}
   				\def \n {15}
   				\def \radius {1.2}
   				\draw circle(\radius)
   				foreach\s in{1,...,\n}{
   					(-360/\n*\s:-\radius)circle(.4pt)circle(.8pt)circle(1.2pt)
   					node[anchor=-360/\n*\s](\s){}
   				};
   				\draw[very thick](1) to [out=-15,in=-120] (5);
   				
   				\draw[very thick](7) to [out=-180,in=100] (11);
   			\end{tikzpicture}
   			\caption*{$[Y_{2}^{(3)}]$}
     		\end{subfigure}
     	\begin{subfigure}[h]{0.29\textwidth}
	\centering
   			\begin{tikzpicture}
   				\def \n {15}
   				\def \radius {1.2}
   				\draw circle(\radius)
   				foreach\s in{1,...,\n}{
   					(-360/\n*\s:-\radius)circle(.4pt)circle(.8pt)circle(1.2pt)
   					node[anchor=-360/\n*\s](\s){}
   				};
   				\draw[very thick](1) to [out=-15,in=-120] (5);
   				
   				\draw[very thick](8) to [out=160,in=80] (12);
   			\end{tikzpicture}
   			\caption*{$[Y_{3}^{(3)}]$}
     		\end{subfigure}
     		\caption{The 3 cyclically distinct weak separation graphs for $n=15, k=3$. Each corresponds to the indicated cyclic class $[Y_\alpha^{(3)}]$ of extremal Yangian invariants containing the element $Y_\alpha^{(3)}$ given in~(\ref{k3Yangian1}) or~(\ref{k3Yangian2}). In each graph the interior of the circle is cut into three disjoint regions, and each disconnected component of the corresponding on-shell diagram (which we omit for the sake of clarity) would lie entirely within a single region (as in Fig~\ref{Y3k}).}\label{k=3}
\end{figure}
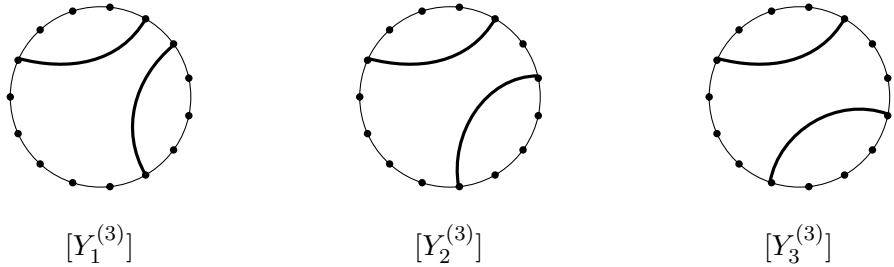

\acknowledgments

We are grateful to A.~Schreiber for collaboration on~\cite{MSSV} which inspired this project and for finding~\cite{OEIS} based on the first few entries of Tab.~\ref{tableofnumbers}, and to J.~Bourjaily for helpful correspondence. This work was supported in part by the US Department of Energy under contract {DE}-{SC}0010010 Task A (MS, AV), and by Simons Investigator Award \#376208 (JM, AV).

\appendix

\section{Proof of the Theorem}\label{proofappendix}

To clarify the theorem which needs to be proven, we begin with an example. Suppose we partition the integers $\set{1,...,12}$ into four subsets $r_1=\set{1,2,6}, r_2=\set{3,4,5}, r_3=\set{7,8,12}, r_4=\set{9,10,11}$. This particular partition is depicted on the left of Fig.~\ref{figpartition}. For any such partition we assign a corresponding $C$-matrix, constructed according to the rule described under~(\ref{egmat}) except that we don't yet gauge fix any parameters to 1. In our example we would have
 \setcounter{MaxMatrixCols}{12}
\begin{equation}\label{egthrm}
 C=
 \begin{pmatrix}
 \alpha_1&\alpha_2&0&0&0&\alpha_6&0&0&0&0&0&0\\
 0&0&\alpha_3&\alpha_4&\alpha_5&0&0&0&0&0&0&0\\
 0&0&0&0&0&0&\alpha_7&\alpha_8&0&0&0&\alpha_{12}\\
 0&0&0&0&0&0&0&0&\alpha_9&\alpha_{10}&\alpha_{11}&0
 \end{pmatrix}.
\end{equation}
The question to be addressed is whether there exists a non-empty domain for the $\alpha$'s on which which matrix is non-negative. (We could also allow $C$ to be non-positive; a non-positive matrix can be made non-negative by exchanging two rows.) We claim:

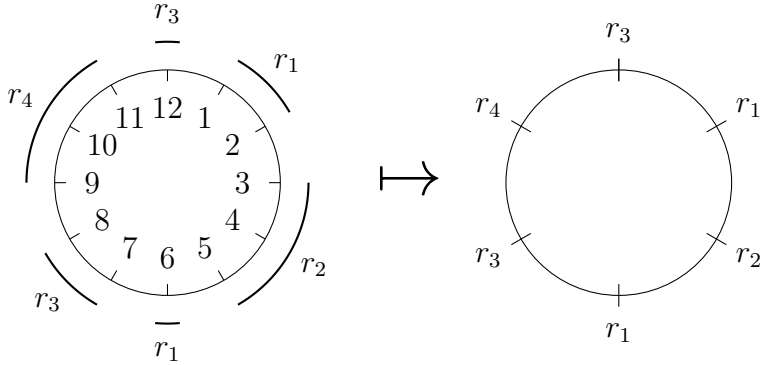
\begin{figure}[t]
\centering
 			\begin{tikzpicture}
 			\def \r {3}
 		\node[circle,draw,minimum size=\r cm] (a) at (0,0) {};
 		\foreach \i [count=\j] in {60,30,...,-270} {
 		    \draw (\i:\r/2) -- (\i:\r/2 -\r/20);
 		    \draw (\i:2*\r/6) node[] {\j};
 		}
 		
 		\draw [thick,domain=30:60] plot ({\r/1.6*cos(\x)}, {\r/1.6*sin(\x)});	
 		\draw (45:3*\r/4) node[] {$r_1$};
 		\draw [thick,domain=0:-60] plot ({\r/1.6*cos(\x)}, {\r/1.6*sin(\x)});	
 		\draw (-30:3*\r/4) node[] {$r_2$};
 		\draw [thick,domain=-85:-95] plot ({\r/1.6*cos(\x)}, {\r/1.6*sin(\x)});	
 		\draw (-90:3*\r/4) node[] {$r_1$};		
 		\draw [thick,domain=-120:-150] plot ({\r/1.6*cos(\x)}, {\r/1.6*sin(\x)});	
 		\draw (-135:3*\r/4) node[] {$r_3$};		
 		\draw [thick,domain=-180:-240] plot ({\r/1.6*cos(\x)}, {\r/1.6*sin(\x)});	
 		\draw (-210:3*\r/4) node[] {$r_4$};	
 		\draw [thick,domain=-265:-275] plot ({\r/1.6*cos(\x)}, {\r/1.6*sin(\x)});	
 		\draw (-270:3*\r/4) node[] {$r_3$};
 		\draw (0:3*\r/2.8) node[] {\Huge{$\mapsto$}};
 		\begin{scope}[shift={(6,0)}]
 					\node[circle,draw,minimum size=\r cm] (a) at (0,0) {};
 					\foreach \i [count=\j] in {90,30,...,-300} {
 					    \draw (\i:\r/2+\r/20) -- (\i:\r/2 -\r/20);
 					}
 					\draw (90:2*\r/3) node[] {$r_3$};
 					\draw (30:2*\r/3) node[] {$r_1$};
 					\draw (-30:2*\r/3) node[] {$r_2$};
 					\draw (-150:2*\r/3) node[] {$r_3$};
 					\draw (-210:2*\r/3) node[] {$r_4$};
 					\draw (-90:2*\r/3) node[] {$r_1$};	
 		\end{scope}		
 			\end{tikzpicture}
 			\caption{The left graph depicts an example partition of the integers $\set{1,...,12}$ into four weakly separated subsets. As the positivity requirement imposes the same conditions on consecutive numbers in the same subset, all of the positivity properties of a partition are contained in its reduced graph, an example of which is depicted on the right.}\label{figpartition}
\end{figure}

\bigskip
\noindent
\textbf{Theorem:} The $C$-matrix associated to a partition of the integers $\set{1,...,n}$ into subsets can be made non-negative if and only if the subsets are all weakly separated.

\bigskip
First notice that the positivity requirement imposes the same restrictions on any set of $\alpha_i$'s which appear consecutively in the same row. For example the domain of $\alpha_3,\alpha_4,\alpha_5$ in (\ref{egthrm}) must clearly be the same. Thus in general we need only study partitions of a ``reduced graph" and its ``contracted matrix". In our example we can simplify our analysis to that of the graph on the right of figure \ref{figpartition} and we may equally well study the corresponding contracted matrix
\begin{equation}
 C'=
 \begin{pmatrix}
 \alpha_1&0&\alpha_6&0&0&0\\
 0&\alpha_3&0&0&0&0\\
 0&0&0&\alpha_8&0&\alpha_{12}\\
 0&0&0&0&\alpha_9&0
 \end{pmatrix}.
\end{equation}
We will only need one lemma to prove the theorem:

\bigskip
\noindent
   \textbf{Lemma:} If a subset $r_i$ occurs more than once around the circle, then the relative sign of the $\alpha$ parameters associated to each of those occurrences is $(-1)^m$ where $m$ is the number of different subsets between those subsets when moving in a clockwise direction, but not counting any occurrences of $r_i$ subsets. This is a necessary but not sufficient condition for ensuring positivity.

\bigskip

For example, between the two $r_1$ elements $1$ and $6$ in Fig.~\ref{figpartition} there is one distinct subset, $r_2$, so the lemma implies that $\alpha_1$ and $\alpha_6$ must have a relative minus sign. If a subset $r_i$ occurs more than once around the circle then we will call the $j^{th}$ clockwise occurrence of it $r_i^j$. To show first that the lemma holds in the specific case of $r_1^1$ and $r_1^2$ in Fig.~\ref{figpartition}, one constructs the following two maximal minors, one involving $r_1^1$ and the other $r_1^2$: first take any selection of particle labels from each of the different subsets between $r_1^1$ and $r_1^2$. In this case there is only one distinct subset between the $r_1$'s, namely $r_2$, and we will take our particle representative of this subset to be particle $3$. To form a maximal minor in this case we will need particle labels from outside of the bounded region, say particles $8\in r_3$ and $10\in r_4$. We now compare the signs of the two so constructed maximal minors:
\begin{align}
  \text{sign}\left(\braket{r_1^1,3,8,10}\right)&=s(r_1^1)s(\alpha_3)s(\alpha_8)s(\alpha_{10})\times\e_{r_1r_2r_3r_4}\label{egsign1}\,,\\
  \text{sign}\left(\braket{3,r_1^2,8,10}\right)&=s(\alpha_3)s(r_1^2)s(\alpha_8)s(\alpha_4)\times \e_{r_2r_1r_3r_4}\label{egsign2}\,,
\end{align}
where $s$ is shorthand for ``sign'' and the $r_1^1$ and $r_1^2$ in the maximal minors denote any particle labels from these subsets. By requiring that~(\ref{egsign1}) and~(\ref{egsign2}) have the same sign (neither is zero, by construction), we determine that entries in $C$ corresponding to particles in $r_1^1$ must differ by a sign from those in $r_1^2$, as our lemma would suggest, e.g.
\begin{equation}
  s(\alpha_1)=-s(\alpha_6)\,.
\end{equation}

With this notation we can prove the lemma in generality. For any two occurrences of the same subset $r_i$ consider the set of subset labels occurring between them in the clockwise direction. Let $m$ denote the number of different subsets within this region, not including $r_i$ subsets. Let $\set{B}$ denote any selection of $m$ particle labels from this region all of which are from different subsets, but not including $r_i$ type particles. In order to form a maximal minor we may need more particle labels from outside of the bounded region. Call any such selection $\set{W}$. Then compare the \textit{signs} of the two maximal minors (here $r_i^j$ refers to any selection of a particle label from the $j^{th}$ occurrence of subset $r_i$):
\begin{equation}
  \braket{r_i^k,\set{B},\set{W}}\quad ,\quad \braket{\set{B},r_i^l,\set{W}}\,.
\end{equation}
To move from one to the other we must pass an $r_i$ through $\set{B}$ which contains $m$ elements. This corresponds to shifting an index on the Levi-Civita symbol $m$ places, thus indicating that $s(r_i^k)=(-1)^ms(r_i^l)$. This proves the lemma.

\bigskip

We can now prove that weak separation implies positivity. We will do so inductively. By our freedom in choosing the $\alpha$'s there is always at least one cyclically ordered maximal minor which is positive. Starting with any cyclically ordered maximal minor we can ``mutate" to generate another cyclically ordered maximal minor in the following manner. If a subset $r_i$ occurs more than once around the circle then given a maximal minor replace the one instance of $r_i$ with the other. For example, starting from the first maximal minor in Fig.~\ref{figpartition} $\braket{r_1^1r_2^1r_3^1r_4^1}$ we could mutate to the $\braket{r_2^1r_1^2r_3^1r_4^1}$ maximal minor. Doing this for all repetitions of subsets will generate all possible cyclically ordered maximal minors. We will show that if the subsets are weakly separated then positivity of the first maximal minor implies that its mutated version is also positive.

Now if the subsets are all weakly separated, then any subset label that occurs within the region bounded by $r_i^k$ and $r_i^l$ cannot occur outside of the bounded region. Therefore, any maximal minor \textit{must} involve $m$ particle labels from $m$ distinct subsets inside the bounded region, where $m$ is the number of distinct subsets, not including $r_i$ subsets, in the bounded region. In this case our condition from the lemma $s(r_i^k)=(-1)^ms(r_i^l)$ is sufficient to ensure that if a maximal minor involving $r_i^k$ is positive, then so is its mutated maximal minor $r_i^j \to r_i^l$. The only possible contradiction would be if we could reach the same maximal minor by mutating on two different maximal minors and reaching conflicting requirements on the sign of one of the $r_i^l$'s. This is not possible in the case of weak separation because for a given mutation, the number of elements that an $r_i$ must pass through in a maximal minor is fixed, being equal to the number of distinct non-$r_i$ subsets in the region bounded to the right of the two $r_i$'s. Thus any two sets of mutations to the same maximal minor necessarily involve the same number column interchanges. This proves that weak separation implies positivity. We now show that not weak separation implies not positivity.

\begin{figure}[t]
     \centering
     			\begin{tikzpicture}
     			\def \r {2.2}
     		
     		\draw [dotted,thick,domain=10:80] plot ({\r/1.4*cos(\x)}, {\r/1.4*sin(\x)});	
     		\draw (90:3*\r/4) node[] {$r_A^1$};
     		\draw (0:3*\r/4) node[] {$r_B^1$};
     		\draw [dotted,thick,domain=-10:-80] plot ({\r/1.4*cos(\x)}, {\r/1.4*sin(\x)});
     		\draw (-90:3*\r/4) node[] {$r_A^2$};
     		\draw [dotted,thick,domain=-100:-170] plot ({\r/1.4*cos(\x)}, {\r/1.4*sin(\x)});
     		\draw (180:3*\r/4) node[] {$r_B^2$};
     		\draw [dotted,thick,domain=100:170] plot ({\r/1.4*cos(\x)}, {\r/1.4*sin(\x)});
     		\draw (45:3*\r/3.3) node[] {$\RN{1}$};
     		\draw (-45:3*\r/3.3) node[] {$\RN{2}$};
     		\draw (-135:3*\r/3.3) node[] {$\RN{3}$};
     		\draw (135:3*\r/3.3) node[] {$\RN{4}$};
    	
     			\end{tikzpicture}
     			\caption{A depiction of any graph which is not weakly separated. The dots refer to any distribution of other subset labels.}\label{notweak}
\end{figure}
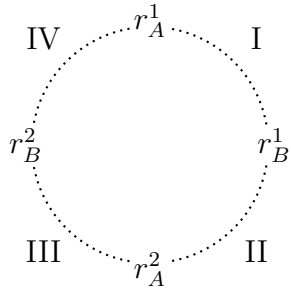

If the subsets are not weakly separated then there exists a subset which occurs more than once, say $r_B^1$ and $r_B^2$, for which there is a subset $r_A^1$ which occurs within the bounded region of these two $r_B$'s with another occurrence $r_A^2$ of it outside of the bounded region. This generic scenario is depicted in Fig.~\ref{notweak}. The difference now is that not every maximal minor must involve $m$ particle labels from the $m$ distinct subsets in the bounded region. The strategy then is to examine a maximal minor which involves $m{-}1$ subsets from the $r_B$ bounded region.

Let $l$ denote the number of distinct subsets in the region $\set{\RN{2}}\union \set{\RN{3}}$ not including $r_A$ type subsets. Let $(\RN{2})$ and $(\RN{3})$ denote any selection of particle labels from $l$ distinct subset in region $\RN{2}$ and $\RN{3}$ not including $r_A$ type particles. To form a complete maximal minor we may require particles from regions $\RN{3}$ and $\RN{4}$. Denote any such generic selection by $(\RN{3})$ and $(\RN{4})$. Then compare the sign of the maximal minors:
\begin{equation}
     		\braket{r_A^1(\RN{1})r_B^1(\RN{2})(\RN{3})(\RN{4})}\quad ,\quad \braket{r_A^1(\RN{1})(\RN{2})(\RN{3})r_B^2(\RN{4})}\,.
\end{equation}
In going from one to the other we must move $r_B$ through $l$ particle labels hence imposing the requirement $s(r_B^1)=(-1)^{l}s(r_B^2)$. However, because there are $l{+}1$ distinct subsets (the $+1$ comes from now including $r_A$ type subsets) between $r_B^1$ and $r_B^2$, our lemma tells us that $s(r_B^1)=(-1)^{l+1}s(r_B^2)$. These are two contradicting requirements. Hence a not weakly separated graph cannot give rise to a positive matrix.\hspace{\fill}$\square$

\bigskip

It is interesting to note that in the $n=5k$ (extremal) case weak separation is the same as planarity. And now that we have shown that weak separation is interchangeable with positivity, we can conclude that the \textit{positivity} of the Grassmannian is in part, or perhaps solely, due to taking the planar limit of a theory.


\begin{thebibliography}{99}

\bibitem{Brink:1976bc}
  L.~Brink, J.~H.~Schwarz and J.~Scherk,
  ``Supersymmetric Yang-Mills Theories,''
  Nucl.\ Phys.\ B {\bf 121}, 77 (1977).

\bibitem{Sohnius:1981sn}
  M.~F.~Sohnius and P.~C.~West,
  ``Conformal Invariance in \texorpdfstring{$\mathcal{N}=4$}{N=4} Supersymmetric Yang-Mills Theory,''
  Phys.\ Lett.\  {\bf 100B}, 245 (1981).

\bibitem{Drummond:2008vq}
  J.~M.~Drummond, J.~Henn, G.~P.~Korchemsky and E.~Sokatchev,
  ``Dual superconformal symmetry of scattering amplitudes in \texorpdfstring{$\mathcal{N}=4$}{N=4} super-Yang-Mills theory,''
  Nucl.\ Phys.\ B {\bf 828}, 317 (2010)
  [arXiv:0807.1095 [hep-th]].

\bibitem{Drummond:2009fd}
  J.~M.~Drummond, J.~M.~Henn and J.~Plefka,
  ``Yangian symmetry of scattering amplitudes in \texorpdfstring{$\mathcal{N}=4$}{N=4} super Yang-Mills theory,''
  JHEP {\bf 0905}, 046 (2009)
  [arXiv:0902.2987 [hep-th]].

\bibitem{Berkovits:2008ic}
  N.~Berkovits and J.~Maldacena,
  ``Fermionic T-Duality, Dual Superconformal Symmetry, and the Amplitude/Wilson Loop Connection,''
  JHEP {\bf 0809}, 062 (2008)
  [arXiv:0807.3196 [hep-th]].

\bibitem{Beisert:2008iq}
  N.~Beisert, R.~Ricci, A.~A.~Tseytlin and M.~Wolf,
  ``Dual Superconformal Symmetry from $AdS_5 \times S^5$ Superstring Integrability,''
  Phys.\ Rev.\ D {\bf 78}, 126004 (2008)
  [arXiv:0807.3228 [hep-th]].

\bibitem{Beisert:2009cs}
  N.~Beisert,
  ``T-Duality, Dual Conformal Symmetry and Integrability for Strings on $AdS_5 \times S^5$,''
  Fortsch.\ Phys.\  {\bf 57}, 329 (2009)
  [arXiv:0903.0609 [hep-th]].

\bibitem{Mason:2009qx}
  L.~J.~Mason and D.~Skinner,
  ``Dual Superconformal Invariance, Momentum Twistors and Grassmannians,''
  JHEP {\bf 0911}, 045 (2009)
  [arXiv:0909.0250 [hep-th]].

\bibitem{ArkaniHamed:2009vw}
  N.~Arkani-Hamed, F.~Cachazo and C.~Cheung,
  ``The Grassmannian Origin Of Dual Superconformal Invariance,''
  JHEP {\bf 1003}, 036 (2010)
  [arXiv:0909.0483 [hep-th]].

\bibitem{ArkaniHamed:2009sx}
  N.~Arkani-Hamed, J.~Bourjaily, F.~Cachazo and J.~Trnka,
  ``Local Spacetime Physics from the Grassmannian,''
  JHEP {\bf 1101}, 108 (2011)
  [arXiv:0912.3249 [hep-th]].

\bibitem{ArkaniHamed:2009dg}
  N.~Arkani-Hamed, J.~Bourjaily, F.~Cachazo and J.~Trnka,
  ``Unification of Residues and Grassmannian Dualities,''
  JHEP {\bf 1101}, 049 (2011)
  [arXiv:0912.4912 [hep-th]].

\bibitem{Drummond:2010qh}
  J.~M.~Drummond and L.~Ferro,
  ``Yangians, Grassmannians and T-duality,''
  JHEP {\bf 1007}, 027 (2010)
  [arXiv:1001.3348 [hep-th]].

\bibitem{Drummond:2010uq}
  J.~M.~Drummond and L.~Ferro,
  ``The Yangian origin of the Grassmannian integral,''
  JHEP {\bf 1012}, 010 (2010)
  [arXiv:1002.4622 [hep-th]].

\bibitem{Ashok:2010ie}
  S.~K.~Ashok and E.~Dell'Aquila,
  ``On the Classification of Residues of the Grassmannian,''
  JHEP {\bf 1110}, 097 (2011)
  [arXiv:1012.5094 [hep-th]].

\bibitem{ArkaniHamed:2012nw}
  N.~Arkani-Hamed, J.~L.~Bourjaily, F.~Cachazo, A.~B.~Goncharov, A.~Postnikov and J.~Trnka,
  ``Grassmannian Geometry of Scattering Amplitudes,''
  arXiv:1212.5605 [hep-th].

\bibitem{Drummond:2008cr}
  J.~M.~Drummond and J.~M.~Henn,
  ``All tree-level amplitudes in \texorpdfstring{$\mathcal{N}=4$}{N=4} SYM,''
  JHEP {\bf 0904}, 018 (2009)
  [arXiv:0808.2475 [hep-th]].

\bibitem{ArkaniHamed:2010kv}
  N.~Arkani-Hamed, J.~L.~Bourjaily, F.~Cachazo, S.~Caron-Huot and J.~Trnka,
  ``The All-Loop Integrand For Scattering Amplitudes in Planar \texorpdfstring{$\mathcal{N}=4$}{N=4} SYM,''
  JHEP {\bf 1101}, 041 (2011)
  [arXiv:1008.2958 [hep-th]].

\bibitem{Arkani-Hamed:2017tmz}
  N.~Arkani-Hamed, Y.~Bai and T.~Lam,
  ``Positive Geometries and Canonical Forms,''
  JHEP {\bf 1711}, 039 (2017)
  [arXiv:1703.04541 [hep-th]].

\bibitem{harary1953number}
  F.~Harary and G.~E.~Uhlenbeck,
  ``On the Number of Husimi Trees, I,''
  Proc.\ Nat.\ Acad.\ Sci.\ {\bf 39}, no. 4, 315 (1953).

\bibitem{bona2000enumeration}
  M.~B{\'o}na, M.~Bousquet, G.~Labelle and P.~Leroux,
  ``Enumeration of $m$-ary cacti,''
  Adv.\ Appl.\ Math.\ {\bf 24}, 22 (2000)
  [math/9804119].

\bibitem{LZ}
  B.~Leclerc and A.~Zelevinsky,
  ``Quasicommuting families of quantum Pl\"ucker coordinates,''
  Adv.\ Math.\ Sci.\ (Kirillov's seminar), AMS Translations 181, 85 (1998).

\bibitem{OPS}
  S.~Oh, A.~Postnikov and D.~E.~Speyer,
  ``Weak separation and plabic graphs,''
  Proc.\ Lond.\ Math.\ Soc.\ {\bf 110}, no. 3, 721 (2015).

\bibitem{ArkaniHamed:2009dn}
  N.~Arkani-Hamed, F.~Cachazo, C.~Cheung and J.~Kaplan,
  ``A Duality For The S Matrix,''
  JHEP {\bf 1003}, 020 (2010)
  [arXiv:0907.5418 [hep-th]].

\bibitem{Hodges:2009hk}
  A.~Hodges,
  ``Eliminating spurious poles from gauge-theoretic amplitudes,''
  JHEP {\bf 1305}, 135 (2013)
  [arXiv:0905.1473 [hep-th]].

\bibitem{Bourjaily:2012gy}
  J.~L.~Bourjaily,
  ``Positroids, Plabic Graphs, and Scattering Amplitudes in Mathematica,''
  arXiv:1212.6974 [hep-th].

\bibitem{OEIS}
  N.\ J.\ A.\ Sloane, editor,
  The On-Line Encyclopedia of Integer Sequences,
  published electronically at {\tt https://oeis.org},
  sequence A054365.

\bibitem{MSSV}
  J.~Mago, A.~Schreiber, M.~Spradlin and A.~Volovich,
  ``Yangian Invariants and Cluster Adjacency in \texorpdfstring{$\mathcal{N}=4$}{N=4} Yang-Mills,''
  arXiv:1906.10682 [hep-th].

\bibitem{Golden:2019kks}
  J.~Golden, A.~J.~McLeod, M.~Spradlin and A.~Volovich,
  ``The Sklyanin Bracket and Cluster Adjacency at All Multiplicity,''
  JHEP {\bf 1903}, 195 (2019)
  [arXiv:1902.11286 [hep-th]].

\bibitem{Drummond:2017ssj}
  J.~Drummond, J.~Foster and {\" O}.~G{\"u}rdo{\u g}an,
  ``Cluster Adjacency Properties of Scattering Amplitudes in \texorpdfstring{$\mathcal{N}=4$}{N=4} Supersymmetric Yang-Mills Theory,''
  Phys.\ Rev.\ Lett.\  {\bf 120}, no. 16, 161601 (2018)
  [arXiv:1710.10953 [hep-th]].

\end{thebibliography}
\end{document}